%% file: rqcopt_hpc.tex
\documentclass[reprint,superscriptaddress,aps,prb,floatfix]{revtex4-2}
\usepackage[ascii]{inputenc}
\usepackage{amsmath,amssymb,amsfonts,amsthm}
\usepackage{mathtools}
\usepackage{babel}
\usepackage{graphicx}
\usepackage{tikz}
\usepackage{pgfplots}
\usepackage{braket}
\usepackage{caption}
\usepackage{subcaption}
\usepackage{enumerate}
\usepackage{listings}
\usepackage[colorlinks=true,linkcolor=magenta,urlcolor=magenta,citecolor=magenta,breaklinks=true]{hyperref}
\usepackage[capitalise]{cleveref}
\usepackage[nolist,nohyperlinks]{acronym}

\DeclareMathOperator{\ud}{d}
\DeclareMathOperator{\e}{e}
\DeclareMathOperator{\Tr}{Tr}
\DeclareMathOperator{\grad}{grad}
\DeclareMathOperator{\hess}{Hess}

\DeclareMathOperator{\asym}{skew}

\DeclareMathOperator*{\argmin}{argmin}

\DeclarePairedDelimiter\norm{\lVert}{\rVert}
\DeclarePairedDelimiter\inner{\langle}{\rangle}

\newcommand{\R}{\mathbb{R}}
\newcommand{\C}{\mathbb{C}}

\graphicspath{{figures/}}
\makeatletter
\def\input@path{{figures/}}
\makeatother

\pgfplotsset{compat = newest}
\usetikzlibrary{arrows.meta}

\colorlet{gatecolor}{blue!15!white}
\def\gw{0.45}
\def\gh{0.75}
\colorlet{statecolor}{purple!15!white}

\definecolor{mblue}  {rgb}{0.368417, 0.506779, 0.709798}
\definecolor{morange}{rgb}{0.880722, 0.611041, 0.142051}
\definecolor{mgreen} {rgb}{0.560181, 0.691569, 0.194885}
\definecolor{mred}   {rgb}{0.922526, 0.385626, 0.209179}
\definecolor{mpurple}{rgb}{0.647624, 0.37816,  0.614037}
\definecolor{mcyan}  {rgb}{0.363898, 0.618501, 0.782349}

\lstdefinestyle{codestyle} {%
    basicstyle=\ttfamily\small,
    commentstyle=\color{mgreen},
    keywordstyle=\color{blue},
    numberstyle=\tiny\color{gray},
    stringstyle=\color{mred},
    showstringspaces=false,
    mathescape=true,
    tabsize=2
}
\begin{document}

\title{High-Performance Contraction of Quantum Circuits for Riemannian Optimization}

\author{Fabian~Putterer}
\thanks{These authors contributed equally}
\email{fabian.putterer@tum.de}
\affiliation{Technical University of Munich, School of Computation, Information and Technology, Boltzmannstra{\ss}e 3, 85748 Garching, Germany}
\author{Max M.~Zumpe}
\thanks{These authors contributed equally}
\email{maxsanchez9819@gmail.com}
\affiliation{Technical University of Munich, School of Computation, Information and Technology, Boltzmannstra{\ss}e 3, 85748 Garching, Germany}
\author{Isabel~Nha~Minh~Le}
\email{isabel.le@tum.de}
\affiliation{Technical University of Munich, School of Computation, Information and Technology, Boltzmannstra{\ss}e 3, 85748 Garching, Germany}
\affiliation{Munich Center for Quantum Science and Technology (MCQST), Schellingstra{\ss}e 4, 80799 Munich, Germany}
\author{Qunsheng~Huang}
\email{keefe.huang@tum.de}
\affiliation{Technical University of Munich, School of Computation, Information and Technology, Boltzmannstra{\ss}e 3, 85748 Garching, Germany}
\affiliation{Munich Center for Quantum Science and Technology (MCQST), Schellingstra{\ss}e 4, 80799 Munich, Germany}
\author{Christian~B.~Mendl}
\email{christian.mendl@tum.de}
\affiliation{Technical University of Munich, School of Computation, Information and Technology, Boltzmannstra{\ss}e 3, 85748 Garching, Germany}
\affiliation{Munich Center for Quantum Science and Technology (MCQST), Schellingstra{\ss}e 4, 80799 Munich, Germany}
\affiliation{Technical University of Munich, Institute for Advanced Study, Lichtenbergstra{\ss}e 2a, 85748 Garching, Germany}

\date{\today}

\begin{abstract}
This work focuses on optimizing the gates of a quantum circuit with a given topology to approximate the unitary time evolution governed by a Hamiltonian. Recognizing that unitary matrices form a mathematical manifold, we employ Riemannian optimization methods -- specifically the Riemannian trust-region algorithm -- which involves second derivative calculations with respect to the gates. Our key technical contribution is a matrix-free algorithmic framework that avoids the explicit construction and storage of large unitary matrices acting on the whole Hilbert space. Instead, we evaluate all quantities as sums over state vectors, assuming that these vectors can be stored in memory. We develop HPC-optimized kernels for applying gates to state vectors and for the gradient and Hessian computation. Further improvements are achieved by exploiting sparsity structures due to Hamiltonian conservation laws, such as parity conservation, and lattice translation invariance. We benchmark our implementation on the Fermi-Hubbard model with up to 16 sites, demonstrating a nearly linear parallelization speed-up with up to 112 CPU threads. Finally, we compare our implementation with an alternative matrix product operator-based approach.
\end{abstract}

\maketitle

\begin{acronym}
    \acro{MPO}{matrix product operator}
    \acro{MPS}{matrix product state}
\end{acronym}

\section{Introduction}

Simulating the unitary time evolution of a quantum system is a natural task for quantum computing~\cite{Lloyd1996, Zalka1998}, with applications to studying out-of-equilibrium behavior of (strongly correlated) systems and as a building block of many quantum algorithms like quantum phase estimation~\cite{NielsenChuang}, the HHL algorithm~\cite{HHL2009} or QETU~\cite{Dong2022}. 
Methods for implementing the Hamiltonian time evolution as a quantum circuit include Trotterization~\cite{Childs2021} and the quantum signal processing framework~\cite{Low2017, 
Low2019, Martyn2021, Motlagh2024}.
Trotterization does not require any auxiliary qubits and allows for reaching long simulation times by repeatedly applying the Trotter circuit, resulting in a linear circuit depth scaling with simulation time.
Still, this procedure can lead to relatively deep circuits. 
Therefore, recent research has focused on improving the approximation accuracy and/or reducing the circuit depth by optimizing the circuit gates, also referred to as quantum circuit compilation~\cite{ge2024quantum}.

In practice, there are several variants of how to approach this task: through an explicit parameterization of quantum gates~\cite{Mansuroglu2023a, Mansuroglu2023b, Tepaske2022, McKeever2023, mc2024towards}, or by updating quantum gates as unitary matrices forming the quantum circuit~\cite{Kotil2024, Le2025riemannian, causer2024scalable, gibbs2024deep, anselme2024combining, rogerson2024quantum, guo2025efficient}. The latter case can be implemented via Riemannian optimization techniques~\cite{luchnikov2021riemannian, Kotil2024, Le2025riemannian, Hauru2021, wiersema2023optimizing, godinez2024riemannian, rogerson2024quantum,  guo2025efficient} by exploiting the mathematical manifold structure of unitary matrices, and is the approach followed in the present work.

Another algorithmic aspect is whether to (i) employ \acp{MPO} and/or \acp{MPS} for representing the target unitary operator and quantum states with controllable accuracy; or (ii) store global (on the full Hilbert space) matrices and/or vectors in computer memory.
Approach~(i) avoids the curse of dimensionality with respect to the quantum Hilbert space dimension but incurs the problem of exponentially increasing virtual bond dimensions with simulated evolution time. Conversely, approach~(ii) is only feasible for relatively small system sizes. However, if the simulated Hamiltonian is translationally invariant, we can extend a circuit optimized for a small system to larger ones by repeating the gates~\cite{Mansuroglu2023a, Kotil2024}. Another advantage of (ii) is its conceptual simplicity and potentially faster runtimes for small system sizes compared to (i).

In this work, we focus on improving the implementation of approach~(ii) via a matrix-free formulation.
As a result, only state vectors -- instead of large unitary matrices -- need to be kept in memory. We present the detailed steps and parallelization strategies for combining this approach with Riemannian quantum circuit optimization based on the trust-region algorithm, which requires second derivatives with respect to the circuit gates. We evaluate our method for the spinless and spinful Fermi-Hubbard models on one-dimensional lattices, reaching up to 16 lattice sites for the spinless model. The parity conservation of these models suggests a natural parity-conserving gate sparsity structure, which we exploit to improve the computational runtime. Finally, we compare our implementation against an alternative method based on \acp{MPO}~\cite{Le2025riemannian} with respect to gradient evaluation time and convergence.

\section{Problem statement and setup}

\subsection{Quantum circuits}

\begin{figure}[!ht]
\centering
\input{gfx/generic_quantum_circuit.tikz}
\caption{Example of a generic quantum circuit consisting of two-qubit gates. The input state vector $\ket{\psi}$ appears on the right, and the gates are applied from right to left.}
\label{fig:generic_quantum_circuit}
\end{figure}
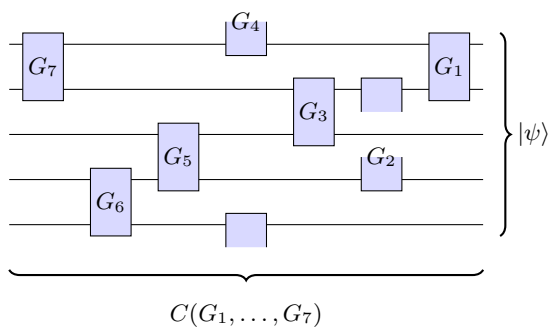

In this paper, we consider qubit quantum circuits for concreteness, but the method can be straightforwardly generalized to ``qudits'' as well. 
We regard a quantum circuit as a sequence of two-qubit gates, which is motivated by the fact that we do not consider any particular hardware-native gate set and that single-qubit gates can be absorbed into the two-qubit gates. 
An example of the considered quantum circuit layout is illustrated in \cref{fig:generic_quantum_circuit}.
We adhere to the mathematical convention of matrix chain ordering from right to left in the circuit drawings, i.e., the gate applied first to the input state vector $\ket{\psi}$ appears on the right. 

$\mathcal{U}(m)$ denotes the set of unitary $m \times m$ matrices, i.e.,
\begin{equation}
\mathcal{U}(m) = \{ V \in \C^{m \times m} \,\vert\, V^{\dagger} V = I_m \},
\end{equation}
with $I_m$ being the $m \times m$ identity matrix.

Formally, a quantum circuit consists of a sequence of two-qubit gates
\begin{equation}
(G_{\ell})_{\ell=1,\dots,n}, \quad \text{with} \quad G_{\ell} \in \mathcal{U}(4) \text{ for all } \ell,
\end{equation}
and associated qubit index tuples $(i_{\ell,1}, i_{\ell,2})_{\ell=1,\dots,n}$ designating the qubits which the individual gates act on.
In principle, some of the gates could be applied in parallel, but for the following derivations regarding gradient computation, the discussion becomes simpler from the equivalent sequential viewpoint.
We write $C(G_1, \dots, G_n)$ for the overall unitary transformation the circuit performs, assuming a fixed circuit topology.

\subsection{Objective function and matrix-free evaluation}

We aim to classically optimize the circuit gates to approximate a target unitary $U$ acting on the overall system.
With $G = (G_1, \dots, G_n)$ denoting the list of gates and quantifying the approximation error by the Frobenius norm distance, the objective can be written as:
\begin{equation}
G_{\text{opt}} = \argmin_{G \in \mathcal{U}(4)^{\times n}} \norm*{C(G) - U}_{\mathrm{F}}^2.
\end{equation}
Since both $U$ and $C(G)$ are unitary by construction,
\begin{equation}
\label{eq:opt_distance}
\norm{C(G) - U}_{\mathrm{F}}^2 = 2\Tr[I] - 2 \, \mathrm{Re}\Tr[U^{\dagger} C(G)],
\end{equation}
where $I$ denotes the identity matrix of the same dimension as $U$. Thus, we may equivalently minimize the following target function:
\begin{equation}
\label{eq:f_target}
f: \mathcal{U}(4)^{\times n} \to \R, \quad f(G) = -\mathrm{Re}\Tr[U^{\dagger} C(G)].
\end{equation}

For $k$ qubits, $U \in \C^{2^k \times 2^k}$, which is a steep cost if stored entirely in memory.
To ameliorate the memory issue, we implement a matrix-free realization of the algorithms in this paper. 
Specifically, we assume that a state vector $\ket{\psi} \in \C^{2^k}$ can still be kept in memory, while the action of $U$ is implemented as a ``matrix-free'' application to a state vector. 
With this setup, we evaluate the target function by a summation over computational basis states:
\begin{equation}
\label{eq:f_target_sum_unit_vec}
f(G) = -\mathrm{Re}\Tr[U^{\dagger} C(G)] = -\mathrm{Re} \sum_{j=0}^{2^k-1} \bra{j} U^{\dagger} C(G) \ket{j}.
\end{equation}
The term $C(G) \ket{j}$ is computed by applying the gates sequentially to each input state $\ket{j}$, equivalent to a state\-vector based circuit simulation, and $\bra{j} U^{\dagger} = (U \ket{j})^{\dagger}$ via the matrix-free action of $U$.

\section{Riemannian optimization}
\label{sec:riemannian_optimization}

To be self-contained, we summarize here the employed mathematical framework of Riemannian optimization on the manifold of unitary matrices~\cite{Edelman1998, Absil2008, Hauru2021}, following the notation in Refs.~\cite{Absil2008, Kotil2024}.

As a central insight, $\mathcal{U}(m)$ for a fixed integer $m$ forms a mathematical manifold, and this additional structure can be used for optimization. 
Technically, $\mathcal{U}(m)$ can be interpreted as a Riemannian submanifold of $\C^{m \times m}$ (with metric described below).
We omit the dimension $m$ from $\mathcal{U}(m)$ for notational simplicity in the following.

The \emph{tangent space} at a given $V \in \mathcal{U}$ is parameterized by the set of complex anti-Hermitian matrices \cite{Absil2008}:
\begin{equation}
\label{eq:tangent_space_param}
T_V\mathcal{U} = \left\{ V A: A \in \C^{m \times m}, A^{\dagger} = -A \right\},
\end{equation}
where $A^{\dagger}$ denotes the adjoint of $A$.
By construction, $V^{\dagger} X$ is anti-Hermitian for any $X \in T_V\mathcal{U}$.

We will use the Euclidean metric as Riemannian metric on $T_V\mathcal{U}$, as in Ref.~\cite{Hauru2021}:
\begin{equation}
\inner{\cdot, \cdot}_V: T_V\mathcal{U} \times T_V\mathcal{U} \to \R, \quad \inner{X, Y}_V = \Tr[X^{\dagger} Y].
\end{equation}
The trace is real-valued since $\Tr[X^{\dagger} Y] = \Tr[(V^\dagger X)^{\dagger} (V^\dagger Y)]$ and $V^\dagger X$, $V^\dagger Y$ are anti-Hermitian.

When viewing $\mathcal{U}$ as embedded into $\C^{m \times m}$, the corresponding projection onto the tangent space at $V \in \mathcal{U}$ reads:
\begin{equation}
\label{eq:proj_tangent}
P_V X = V \asym(V^{\dagger} X),
\end{equation}
with $\asym(A) = \frac{1}{2} (A - A^{\dagger})$ the anti-Hermitian part of a matrix~\cite{Absil2008, Hauru2021}.

We define the \emph{gradient} of a smooth function $f: \C \to \R$ at a point $z = x + i y$ with $x, y \in \R$ as a complex number constructed from the derivatives with respect to the real and imaginary components of $z$:
\begin{equation}
\label{eq:def_grad}
\grad f(z) = \partial_x f(z) + i \partial_y f(z).
\end{equation}
We apply this definition entry-wise for functions depending on several complex numbers, e.g., $f: \C^m \to \R$ or $f: \C^{m \times m} \to \R$. For the latter case, this convention for the gradient implies that 
\begin{equation}
\frac{\ud f(G + \epsilon X)}{\ud \epsilon} = \mathrm{Re}\Tr\big[(\grad f(G))^{\dagger} X\big]
\end{equation}
for all $G, X \in \C^{m \times m}$ and $\epsilon \in \R$~\cite{Hauru2021}.

Regarding a circuit gate as a general matrix corresponds to interpreting the to-be optimized target function $f: \mathcal{U} \to \R$ as a restriction of a function $\bar{f}$ defined on $\C^{m \times m}$. In this case, the gradient vector of $f$ results from projecting the gradient vector of $\bar{f}$ onto the tangent space:
\begin{equation}
\label{eq:grad_proj}
\grad f(V) = P_V \grad \bar{f}(V).
\end{equation}

Defining second derivatives and Hessian matrices requires some additional concepts.
Let $\mathfrak{X}(\mathcal{U})$ denote the set of smooth vector fields on $\mathcal{U}$, following the notation of Ref.~\cite{Absil2008}.
We will use the unique \emph{Riemannian (Levi-Civita) connection} $\nabla$, which is formally defined as a map
\begin{equation}
\nabla: \mathfrak{X}(\mathcal{U}) \times \mathfrak{X}(\mathcal{U}) \to \mathfrak{X}(\mathcal{U}), \quad (\eta, \xi) \mapsto \nabla_{\eta} \xi
\end{equation}
which is symmetric and compatible with the Riemannian metric. Intuitively, $\nabla_{\eta}$ is the derivative of a vector field in direction $\eta$.

As before, we interpret $\mathcal{U}$ as a Riemannian submanifold of $\C^{m \times m}$. Let $\xi \in \mathfrak{X}(\mathcal{U})$ be a vector field.
Then the derivative of $\xi$ in gradient direction $X \in T_V \mathcal{U}$ ($V \in \mathcal{U}$) is given by~\cite[Eq.~(5.15)]{Absil2008}
\begin{equation}
\label{eq:vector_field_derivative}
\nabla_X \xi = P_V (D\xi(V)[X]),
\end{equation}
where $D\xi(V)[X]$ is the (Euclidean) gradient of $\xi$ in direction $X$ at point $V$.

A \emph{retraction} on $\mathcal{U}$~\cite[chapter~4]{Absil2008} is a mapping from the tangent bundle of the unitary matrix manifold into the manifold. We have found it convenient to use the polar decomposition ($V \in \mathcal{U}$) as a retraction:
\begin{equation}
\label{eq:retraction_qr}
R: T\mathcal{U} \to \mathcal{U}, \quad R_V(\xi) = q_{\text{polar}}(V + \xi),
\end{equation}
where $q_{\text{polar}}(A)$ denotes the unitary matrix $Q \in \mathcal{U}$ from the polar decomposition of $A \in \C^{m \times m}$ as $A = Q P$, with $P$ a Hermitian positive semi-definite matrix of the same size as $A$.

In our numerical calculations, we will simultaneously update all the quantum gates $G_1, \dots, G_n$ in an optimization step. 
Matching this procedure with the mathematical formalism requires a generalization to target functions depending on several unitary matrices:
\begin{equation}
\label{eq:f_target_abstract}
f: \underbrace{\mathcal{U} \times \cdots \times \mathcal{U}}_{n \text{ terms}} \to \R.
\end{equation}
Formally, $f$ is a function from the product manifold $\mathcal{U}^{\times n}$ to the real numbers. 
The corresponding tangent space is the direct sum of the individual tangent spaces, cf.~\cite{Hauru2021}.
In practice, the overall gradient vector is thus a concatenation of the individual gradient vectors in \cref{eq:grad_proj}, and the overall retraction results from applying the retraction in \cref{eq:retraction_qr} to the individual isometries and tangent vectors. 

For minimizing $f$ in \cref{eq:f_target_abstract}, we will use the Riemannian trust-region algorithm~\cite[chapter~7]{Absil2008}.
The central idea consists of a quadratic approximation of the target function in the neighborhood of a point $G \in \mathcal{U}^{\times n}$:
\begin{equation}
\hat{m}_G(X) = f(G) + \inner{\grad f(G), X} + \frac{1}{2} \inner{\hess f(G)[X], X}
\end{equation}
for $X \in T_G \mathcal{U}^{\times n}$, with the Riemannian Hessian
\begin{equation}
\label{eq:hessian_def}
\hess f(G)[X] = \nabla_X \grad f(G).
\end{equation}
The specific details for computing the gradient and Hessian will be discussed in the next section. The remaining ingredients are the same as in Ref.~\cite{Kotil2024}: we implement the Riemannian trust-region algorithm \cite[Algorithm~10]{Absil2008}, using the truncated conjugate-gradient method for the trust-region subproblem, see \cite[Algorithm~11]{Absil2008} and Ref.~\cite{Steihaug1983}.

\section{Optimized circuit contraction and derivative computation}
\label{sec:circuit_contraction_derivative_computation}

\begin{figure}[!ht]
\centering
\input{gfx/target_network.tikz}
\caption{Tensor network representation of a target function summand (before taking the real part), see \cref{eq:f_target_sum_unit_vec}.}
\label{fig:target_network}
\end{figure}
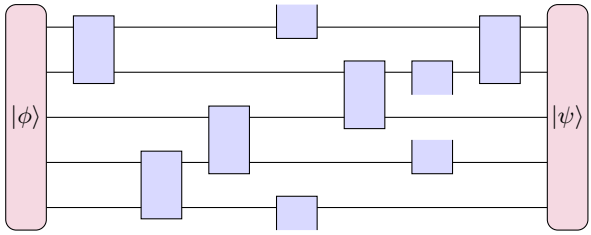

According to \cref{eq:f_target_sum_unit_vec}, the task of evaluating the target function and computing gradients is reduced to evaluating tensor networks of the form shown in \cref{fig:target_network}, each representing one of the summands in \cref{eq:f_target_sum_unit_vec}.
In our case, $\ket{\psi} = \ket{j}$ and $\ket{\phi} = (U \ket{j})^*$ for the $j$-th summand, where we adhere to the convention that $\ket{\phi}$ is not complex-conjugated in the tensor diagram. The following calculations apply to general state vectors $\ket{\psi}, \ket{\phi} \in \C^{2^k}$.

\subsection{Gradient and Hessian tensor diagrams}

The derivations in this subsection are based on the insight that we can treat the to-be optimized gates $\{G_{\ell}\}$ as general complex matrices in $\C^{m \times m}$ at first and accordingly interpret the target function as restriction and the real part of a function $\tilde{f}: (\C^{m \times m})^{\times n} \to \C$.
We perform the gradient and Hessian computation for $\tilde{f}$ and then account for the unitary submanifold structure by appropriate projections of the gradients, as explained in \cref{sec:riemannian_optimization}.
$\tilde{f}$ will be holomorphic, and accordingly, we take complex derivatives in this subsection.

\begin{figure}[!ht]
\centering
\input{gfx/network_gradient_hole.tikz}
\caption{Tensor network representation of the derivative with respect to the entries of a gate $G_{\ell}$ (interpreted as a general matrix). The open tensor legs due to the missing gate define the gradient with respect to $G_{\ell}$.}
\label{fig:network_gradient_hole}
\end{figure}
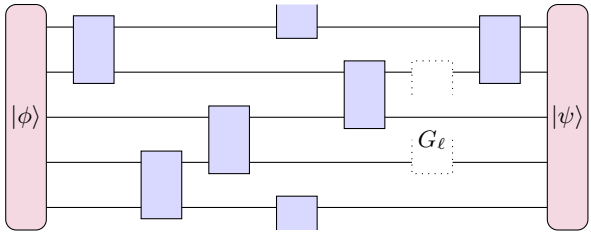

The gradient with respect to the entries of a gate $G_{\ell}$ corresponds to the identical diagram with $G_{\ell}$ omitted, as described in \cite{Kotil2024}, illustrated in \cref{fig:network_gradient_hole}. We identify such omissions in diagrams with labeled, dotted outlines and refer to them as gate holes.

The overall gradient consists of the concatenated gradients with respect to all gates.
Thus, a natural question is how to organize these computations to re-use intermediate results, i.e., how to cache intermediate quantities efficiently.
We will study this question in the following section.

\begin{figure}[!ht]
\centering
\input{gfx/network_hessian_holes.tikz}
\caption{Tensor network representation of the second derivative with respect to the entries of two gates.}
\label{fig:network_hessian_holes}
\end{figure}
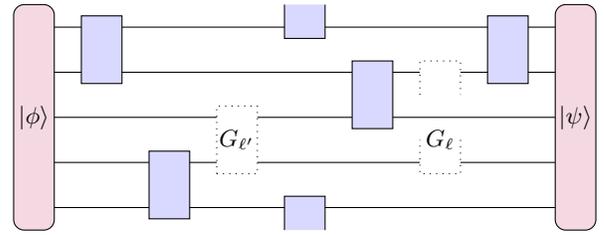

The Hessian matrix assumes a block structure.
The blocks have size $m^2 \times m^2$ and correspond to the derivatives with respect to any pair of gates, denoted $G_{\ell}$ and $G_{\ell'}$.
Note that the target function (and correspondingly each summand) is a \emph{linear} function of each individual gate.
In particular, the second derivative with respect to the matrix entries of a single gate vanishes.
Due to symmetry, we only need to consider the upper or lower block-triangular part, thus w.l.o.g.\ $\ell < \ell'$.
In the diagrammatic picture, this derivative corresponds to two gate holes in the network, as illustrated in \cref{fig:network_hessian_holes}.

\subsection{Efficient gradient and Hessian computation using caching/backpropagation techniques}
Naively, the tensor diagrams in \cref{fig:network_gradient_hole} for all $\ell = 1, \dots, n$ can be evaluated by performing a gate-wise contraction of the left and right sides of the hole, starting from the respective state vectors at the boundaries. Doing so separately for each $\ell$ will involve redundant computations over various gate holes.

To avoid these redundancies, we start from the insight that the intermediate quantum states resulting from the sequential gate application to $\ket{\psi}$ can be cached.
The resulting algorithm is equivalent to backpropagation used in artificial neural networks.
In essence, each gate multiplication is interpreted as a layer of a feed-forward neural network.
The gradients with respect to the gates can be obtained by a forward and backward pass through the network.

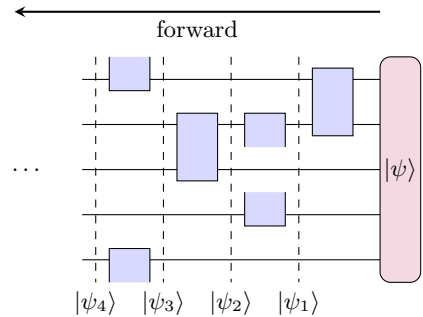
\begin{figure}[!ht]
\centering
\input{gfx/forward_pass.tikz}
\caption{Forward pass through the gate network, caching (storing) the intermediate vectors $\ket{\psi_1}, \ket{\psi_2}, \dots$.}
\label{fig:forward_pass}
\end{figure}

The forward pass consists of the usual sequential gate applications (without holes), storing all intermediate state vectors $\ket{\psi_1}, \ket{\psi_2}, \dots$ as shown in \cref{fig:forward_pass}.
These state vectors correspond to the caching mechanism and will be reused for computing the gradient on each gate hole.
The detailed procedure for applying a gate to a state vector will be described below in \cref{sec:statevector-operations}.

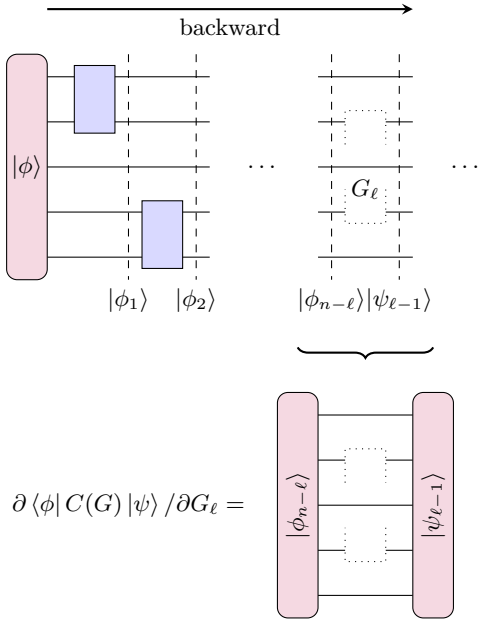
\begin{figure}[!ht]
\centering
\input{gfx/backward_pass.tikz}
\caption{Backward pass and gate gradient computation for one summand of the target function.}
\label{fig:backward_pass_gate_gradient}
\end{figure}

The backward pass proceeds in reverse order, with $\ket{\phi}$ as the input vector and applying the transposed gates (as a consequence of the chain rule), see \cref{fig:backward_pass_gate_gradient}.
$\ket{\phi}$ plays the role of the gradient of the target function with respect to the circuit output.
The gradient with respect to gate $G_{\ell}$ is a hole in the network in the diagrammatic representation, and can be computed by ``sandwiching'' the intermediate backward-pass state $\ket{\phi_{n-\ell}}$ with $\ket{\psi_{\ell-1}}$ from the forward pass, as shown in the lower half of \cref{fig:backward_pass_gate_gradient}.
Note that the gradients can be calculated ``on the fly'' during the backward pass.

The overall gradient computation thus requires $\mathcal{O}(n)$ gate applications to a state vector (instead of $\mathcal{O}(n^2)$ for the naive approach).

Our next goal is to compute the entire Hessian matrix.
Alternatively, one could evaluate Hessian-vector products at a cost scaling linearly with the number of gates.
Although this might appear more efficient, it turns out that the first option is computationally advantageous in our case, given the subsequent optimization algorithm in our setting.
For completeness, we describe the Hessian-vector product computation in \cref{sec:Hessian-vector_product}.

As for the gradient, it suffices to consider the Hessian computation for an individual summand in \cref{eq:f_target_sum_unit_vec}.

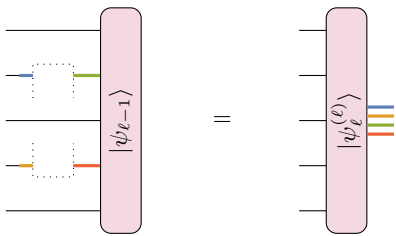
\begin{figure}[!ht]
\centering
\input{gfx/hole_application.tikz}
\caption{Applying a gate hole to a state vector. The open legs corresponding to the missing gate are identified by colors and grouped into one logical leg of dimension $m^2$ on the right (state vector array index).}
\label{fig:hole_application}
\end{figure}

\begin{figure}[!ht]
\centering
\input{gfx/hessian_pass.tikz}
\caption{Second forward pass for computing blocks of the Hessian matrix. The second derivative with respect to $G_{\ell'}$ is evaluated by sandwiching the forward-propagated hole originating from the first derivative with respect to another gate $G_{\ell}$.}
\label{fig:hessian_pass_second_gradient}
\end{figure}
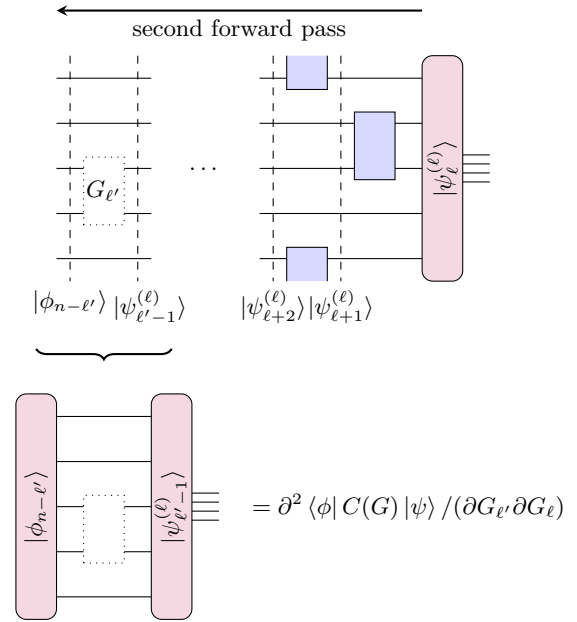

To evaluate the diagram in \cref{fig:network_hessian_holes} for all pairs $1 \le \ell < \ell' \le n$ while avoiding re-computations, we additionally store the vectors $\ket{\phi_1}, \ket{\phi_2}, \dots$ from the gradient backward pass, besides the state vectors $\ket{\psi_1}, \ket{\psi_2}, \dots$ from the forward pass.
Then, for each $\ell = 1, \dots, n-1$, we perform another forward pass starting from layer $\ell$ and a special state vector array denoted $\ket{\psi^{(\ell)}_\ell}$.
This array is the result of applying the hole of gate $G_{\ell}$ to $\ket{\psi_{\ell-1}}$, as illustrated in \cref{fig:hole_application}.
It has an additional tensor leg (array index) of dimension $m^2$, corresponding to the entries of the missing gate $G_{\ell}$.
From an alternative viewpoint, these are the first derivatives with respect to the $m^2$ entries of $G_{\ell}$.
The isolated wire segments on the left of the gate hole can be interpreted as identity maps.
Thus, the operation is equivalent to the Kronecker product between the state vector and an $m \times m$ identity matrix, followed by a permutation of tensor legs.
The state vector array $\ket{\psi^{(\ell)}_\ell}$ is then propagated forward through the circuit by successively applying the gates $G_{\ell+1}, G_{\ell+2}, \dots$ to it, as for a conventional state vector, see \cref{fig:hessian_pass_second_gradient}.
In each layer $\ell' = \ell+1, \dots, n$, the hole corresponding to $G_{\ell'}$ is sandwiched between $\ket{\phi_{n-\ell'}}$ and $\ket{\psi^{(\ell)}_{\ell'-1}}$ as visualized in the lower half of \cref{fig:hessian_pass_second_gradient}, analogous to the backward pass for the gradient computation. This computes the second derivative for every encountered layer $\ell'$ relative to layer $\ell$ in a single forward pass.

In our implementation, we interleave the entries of the logical states in a state vector array, i.e., the four right-pointing tensor legs of $\ket{\psi^{(\ell)}_\ell}$ are the fastest-running indices. This enables computationally advantageous vectorization when applying subsequent gates, since the mathematical operations on blocks of $m^2$ entries (which are adjacent in memory) can be performed in parallel.

\subsection{Details of the state vector simulation}
\label{sec:statevector-operations}

The framework described in the preceding sections can finally be implemented based on the following elementary building blocks: applying a two-qubit gate to a state vector, sandwiching a gate hole between two state vectors, and performing the ``hole'' application in \cref{fig:hole_application}.
We explain these building blocks in the following subsections.
A single state vector and a state vector array are always stored as a contiguous block with complex entries in memory.

\subsubsection{Gate application}

\begin{figure}[!ht]
\centering
\begin{subfigure}[t]{0.5\columnwidth}
\input{gfx/gate_application_neighboring.tikz}%
\caption{neighboring wires}
\label{fig:gate_application_neighboring}
\end{subfigure}%
\begin{subfigure}[t]{0.5\columnwidth}
\input{gfx/gate_application_general.tikz}
\caption{general case}
\label{fig:gate_application_general}
\end{subfigure}
\caption{Applying a two-qubit gate to a state vector. (a) shows the special case in which the gate acts on neighboring wires, and (b) the general case. Multiple qubit wires have been combined into single tensor legs.}
\label{fig:gate_application}
\end{figure}
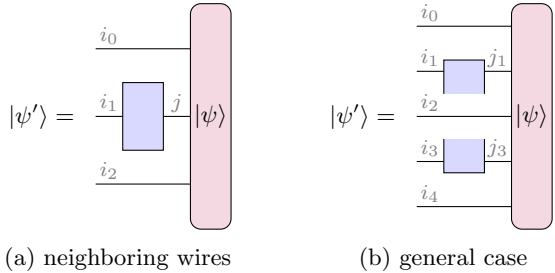

A gate application can be interpreted as a tensor network contraction, as visualized in \cref{fig:gate_application}.
We treat the special case where the gate acts on two neighboring qubits separately, since the indexing is simpler in this case.
The qubit wires not affected by the gate have been merged into single tensor legs as far as possible.
Note that some of the legs can have (dummy) dimension $0$.
By convention, the leg indexed by $i_0$ corresponds to the most significant dimension (slowest running index), and we use zero-based indexing.
The contiguous index of a state vector entry can be computed by lexicographical enumeration.
For example, in the context of \cref{fig:gate_application_neighboring}, assuming that the leg indexed by $i_2$ has dimension $n$ and the leg indexed by $j$ dimension $m$, then
\begin{equation}
\ket{\psi}_{i_0, j, i_2} = \psi[(i_0 \cdot m + j) \cdot n + i_2],
\end{equation}
where $\psi[i]$ is the $i$-th entry of the state vector memory block.

Regarding the case of neighboring wires, we denote the gate entries as $g_{i,j}$.
The entries of the output state vector $\ket{\psi'}$ in \cref{fig:gate_application_neighboring} are then given by
\begin{equation}
\label{eq:gate_application_neighboring}
\ket{\psi'}_{i_0, i_1, i_2} = \sum_{j=0}^3 g_{i_1, j} \cdot \ket{\psi}_{i_0, j, i_2} \quad \forall i_0, i_1, i_2.
\end{equation}

In the general case, we interpret the input and output state vectors as tensors of degree $5$ and the two-qubit gate as a tensor of degree $4$, as shown in \cref{fig:gate_application_general}.
The analogous formula to obtain the output state is then
\begin{equation}
\ket{\psi'}_{i_0, i_1, i_2, i_3, i_4} = \sum_{j_1, j_3=0}^1 g_{i_1, i_3, j_1, j_3} \cdot \ket{\psi}_{i_0, j_1, i_2, j_3, i_4}
\end{equation}
for all $i_0, \dots, i_4$.

A straightforward implementation of \cref{eq:gate_application_neighboring}, assuming that the leg indexed by $i_0$ has dimension $p$ and that the gate is an $m \times m$ matrix, reads:
\begin{lstlisting}[language=C, label=lst:gate-application-pseudo-code]
$\psi' = (0, \dots, 0)$  // initialize as zero vector
for ($i_0 \in \{ 0, \dots, p - 1 \}$)
{
  for ($i_1 \in \{ 0, \dots, m - 1 \}$)
  {
    for ($j \in \{ 0, \dots, m - 1 \}$)
    {
      for ($i_2 \in \{ 0, \dots, n - 1 \}$)
      {
        $\psi'[(i_0 \cdot m + i_1) \cdot n + i_2]$ +=
          $g[i_1 \cdot m + j]$ * $\psi[(i_0 \cdot m + j) \cdot n + i_2]$
      }
    }
  }
}
\end{lstlisting}
The fastest-running index appears in the inner loop.

Alternatively, one can use the loops over the gate entries as the inner ``kernel'' and eventually unroll them. For the case of qubits ($m = 4$), this results in
\begin{lstlisting}[language=C, label=lst:gate-application-pseudo-code-alt]
for ($i_0 \in \{ 0, \dots, p - 1 \}$)
{
  for ($i_2 \in \{ 0, \dots, n - 1 \}$)
  {
    $x_0$ = $\psi[(i_0 \cdot 4 + 0) \cdot n + i_2]$
    $x_1$ = $\psi[(i_0 \cdot 4 + 1) \cdot n + i_2]$
    $x_2$ = $\psi[(i_0 \cdot 4 + 2) \cdot n + i_2]$
    $x_3$ = $\psi[(i_0 \cdot 4 + 3) \cdot n + i_2]$

    // gate multiplication kernel
    $y$ = (0, 0, 0, 0)
    for ($k \in \{ 0, \dots, 3 \}$)
    {
      for ($j \in \{ 0, \dots, 3 \}$)
      {
        $y_k$ += $g[k \cdot 4 + j]$ * $x_j$
      }
    }

    $\psi'[(i_0 \cdot 4 + 0) \cdot n + i_2]$ = $y_0$
    $\psi'[(i_0 \cdot 4 + 1) \cdot n + i_2]$ = $y_1$
    $\psi'[(i_0 \cdot 4 + 2) \cdot n + i_2]$ = $y_2$
    $\psi'[(i_0 \cdot 4 + 3) \cdot n + i_2]$ = $y_3$
  }
}
\end{lstlisting}

As the operations performed are first reading, then performing simple arithmetic operations, recombining, and finally writing data, they are limited by the performance of the memory architecture, not the arithmetic units. Therefore, the algorithm should read data sequentially to exploit cache locality. While the application process described above reads state vector entries in locations separated by a stride of a power of 2, it still does so in four individual, locally sequential streams. This means that the operation, while not strictly sequential, behaves like a cache-local process and, therefore, has optimal performance regarding memory access.

The general case shown in \cref{fig:gate_application_general} leads to an analogous straightforward generalization of the above code, with a larger nesting depth due to the additional tensor legs.

\subsubsection{Gate gradient as a gate hole}

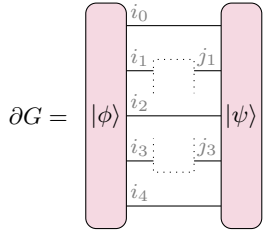
\begin{figure}[!ht]
\centering
\input{gfx/hole_contraction.tikz}
\caption{Tensor network to obtain the gradient w.r.t.\ the missing gate. The qubit wires not acted on by the gate are grouped into the legs indexed by $i_0$, $i_2$, $i_4$.}
\label{fig:hole_contraction}
\end{figure}

As described above, the gradient or Hessian computation entails contracting a tensor network with a gate hole (see bottom parts of \cref{fig:backward_pass_gate_gradient} and \cref{fig:hessian_pass_second_gradient}). This step is summarized in \cref{fig:hole_contraction}, where dimensions are grouped analogously to the gate application.

A pseudocode implementation to perform this contraction for the case of qubits, assuming that the leg dimensions for $i_0$, $i_2$, $i_4$ are $p$, $q$, $n$, respectively, reads:
\begin{lstlisting}[language=C]
$\partial G = 0 \in \C^{4 \times 4}$  // initialize as zero matrix
for ($i_0 \in \{ 0, \dots, p - 1 \}$)
{
  for ($i_2 \in \{ 0, \dots, q - 1 \}$)
  {
    for ($i_4 \in \{ 0, \dots, n - 1 \}$)
    {
      $x_0$ = $\psi[(((i_0 \cdot 2 + 0) \cdot q + i_2) \cdot 2 + 0) \cdot n + i_4]$
      $x_1$ = $\psi[(((i_0 \cdot 2 + 0) \cdot q + i_2) \cdot 2 + 1) \cdot n + i_4]$
      $x_2$ = $\psi[(((i_0 \cdot 2 + 1) \cdot q + i_2) \cdot 2 + 0) \cdot n + i_4]$
      $x_3$ = $\psi[(((i_0 \cdot 2 + 1) \cdot q + i_2) \cdot 2 + 1) \cdot n + i_4]$

      $y_0$ = $\phi[(((i_0 \cdot 2 + 0) \cdot q + i_2) \cdot 2 + 0) \cdot n + i_4]$
      $y_1$ = $\phi[(((i_0 \cdot 2 + 0) \cdot q + i_2) \cdot 2 + 1) \cdot n + i_4]$
      $y_2$ = $\phi[(((i_0 \cdot 2 + 1) \cdot q + i_2) \cdot 2 + 0) \cdot n + i_4]$
      $y_3$ = $\phi[(((i_0 \cdot 2 + 1) \cdot q + i_2) \cdot 2 + 1) \cdot n + i_4]$

      $\partial G$ += $y \otimes x$  // outer product
    }
  }
}
\end{lstlisting}
The inner ``kernel'' is the outer product of the two four-dimensional vectors $x$ and $y$.

For the Hessian computation, $\ket{\psi}$ needs to be substituted by a state vector array, which leads to an additional loop in the implementation over the (inner) array dimension.

The algorithm for the gate gradient is almost equivalent to the gate application, differing only in the direction of data flow. The latter reads a state vector, applies a gate, and writes the result to another state vector. Conversely, the gate gradient computation reads two state vectors and combines them into a gradient. Apart from this, the indexing of the contraction is identical. Consequently, the performance of the gate gradient calculation is expected to be similar to that of the gate application.

\subsubsection{Gate hole application}

Finally, we discuss the gate hole application to a state vector as visualized in \cref{fig:hole_application}.
We interleave the individual state vector entries in the resulting array in our implementation. In other words, a state vector array $(\ket{\psi_1}, \ket{\psi_2}, \dots)$ adheres to the linear memory layout $(\psi_1[0], \psi_2[0], \dots, \psi_1[1], \psi_2[1], \dots)$.

Assuming that the tensor legs that are not affected by the gate hole in \cref{fig:hole_application} have dimensions $p$, $q$, $n$, respectively, the subroutine for the case of qubits is implemented by the following pseudocode:
\begin{lstlisting}[language=C]
for ($i_0 \in \{ 0, \dots, p - 1 \}$)
{
  for ($i_2 \in \{ 0, \dots, q - 1 \}$)
  {
    for ($i_4 \in \{ 0, \dots, n - 1 \}$)
    {
      $x_0$ = $\psi[(((i_0 \cdot 2 + 0) \cdot q + i_2) \cdot 2 + 0) \cdot n + i_4]$
      $x_1$ = $\psi[(((i_0 \cdot 2 + 0) \cdot q + i_2) \cdot 2 + 1) \cdot n + i_4]$
      $x_2$ = $\psi[(((i_0 \cdot 2 + 1) \cdot q + i_2) \cdot 2 + 0) \cdot n + i_4]$
      $x_3$ = $\psi[(((i_0 \cdot 2 + 1) \cdot q + i_2) \cdot 2 + 1) \cdot n + i_4]$

      // equivalent to outer product
      // with 4x4 identity matrix
      // and transpositions
      for ($k \in \{ 0, 1 \}$)
      {
        for ($\ell \in \{ 0, 1 \}$)
        {
          $j$ = $(((((i_0 \cdot 2 + k) \cdot q + i_2) \cdot 2 + \ell) \cdot n$
                        $+ i_4) \cdot 2 + k) \cdot 2 + \ell$
          $\psi'[j \cdot 4 + 0]$ = $x_0$
          $\psi'[j \cdot 4 + 1]$ = $x_1$
          $\psi'[j \cdot 4 + 2]$ = $x_2$
          $\psi'[j \cdot 4 + 3]$ = $x_3$
        }
      }
    }
  }
}
\end{lstlisting}

\subsection{Brick wall circuit layout}

A brick wall layout is a common choice for the circuit topology, appearing, for example, in the TEBD algorithm, and was considered in Ref.~\cite{Kotil2024}, which the present work builds upon. From another viewpoint, the brick wall layout applies as many gates as possible in parallel, which leads to a minimal circuit depth, given the overall number of gates. In the present work, we use a sequential gate order to simplify the mathematical formalism. Nevertheless, this is not a principal restriction, and in this subsection, we demonstrate how to transform the brick wall ansatz (and parallel gates in general) into the present formalism. 

\begin{figure}[!ht]
\centering
\input{gfx/brickwall_to_sequential.tikz}
\caption{Transforming a brick wall layout to a sequential gate sequence.}
\label{fig:brickwall_to_sequential}
\end{figure}
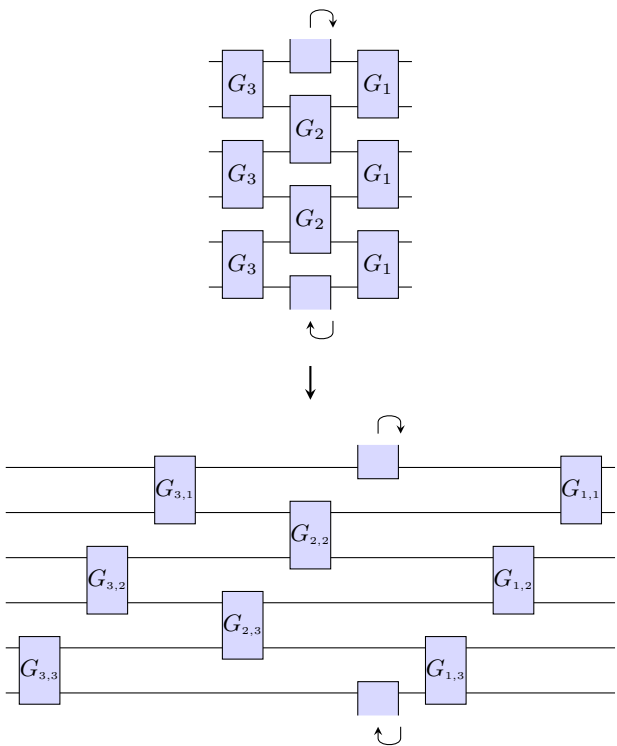

\cref{fig:brickwall_to_sequential} shows a straightforward ``sequentialization'' of a brick wall structure without changing the circuit's logical operation. An additional aspect is the multiple appearances of a gate. For example, in the brick wall circuit, each layer uses copies of the same gate due to translational invariance. For computing derivatives, it is advantageous to list gates in a sequential ordering and formally treat them as separate gates at first. In \cref{fig:brickwall_to_sequential}, the repeated gate $G_{\ell}$ becomes $G_{\ell,i}$ (with same entries as $G_{\ell}$). Computing derivatives then proceeds as described in \cref{sec:circuit_contraction_derivative_computation} above. Finally, to obtain the gradient with respect to $G_{\ell}$, the gradients with respect to the $G_{\ell,i}$ gates are accumulated:
\begin{equation}
\partial G_{\ell} = \sum_i \partial G_{\ell,i}.
\end{equation}
An analogous formula holds for the Hessian matrix.

\subsection{Parity-conserving gate structure}

The Hessian computation can be simplified if the two-qubit gates have certain structures. For instance, when considering fermionic models such as the Hubbard model, we can use the Ansatz that the individual gates inherit the parity structure of the overall model. Explicitly, every gate in the circuit then has the following form:
\begin{equation}
G =
\begin{pmatrix}
g_{11} & 0      & 0      & g_{14}\\
0      & g_{22} & g_{23} & 0     \\
0      & g_{32} & g_{33} & 0     \\
g_{41} & 0      & 0      & g_{44}\\
\end{pmatrix}, 
\label{eq:paritygatematrix}
\end{equation}
with unitary sub-blocks
\begin{equation}
G_1 = \begin{pmatrix} g_{22} & g_{23} \\ g_{32} & g_{33}\end{pmatrix} \in \mathcal{U}(2), \quad G_2 = \begin{pmatrix} g_{11} & g_{14} \\ g_{41} & g_{44}\end{pmatrix} \in \mathcal{U}(2).
\end{equation}
The structure of $G$ results from assigning even parity to the first and last row and column indices, and odd parity to the second and third, and requiring that the gate cannot change parity. The zero elements of the gate matrices can be ignored in the optimization procedure. This does not lead to significant speedups in the forward and backward passes, as applying gates mainly scales with the size of the state vector. However, in the second forward pass, the number of temporary state vectors that need to be propagated to the second gate-hole is reduced by the number of zero elements in the updated matrices. This is a factor of two in the case of parity-conserving gates like \cref{eq:paritygatematrix}. Note that further restrictions could be imposed on the gates, such as particle number conservation, which would remove another two parameters.

\subsection{Exploiting translational invariance}

Since the intended use case of our procedure is translationally invariant systems and brick wall topologies, we can exploit this feature to reduce the number of gate-hole pairs that need to be computed for the Hessian. Notice that if any two-hole configuration differs from another by a global periodic translation of all lattice sites, then their numerical contribution to the Hessian is the same. We can limit our computation to the unique two-hole combinations under translations.


\section{Resource requirements and computational complexity}

\subsection{Memory considerations for $16$ qubits}

For illustration purposes, we will assume a circuit with 16 qubits. Forming the entire matrix for each layer using the Kronecker product would require $8 \,\text{B} \cdot 2 \cdot (2^{16})^2 = 64 \,\text{GiB}$ per layer for 16~qubits and complex double-precision floating point values. Using the matrix-free method described in this work, the memory footprint can be significantly reduced, as only state vectors and small gate matrices need to be stored. With a 16-qubit system, each state vector has $2^{16} = 65536$ entries and therefore uses exactly 1~MiB. During the contraction, buffers for the left and right state vectors $\ket{\phi_{n-\ell}}$ and $\ket{\psi_{\ell-1}}$ are needed, and 2~MiB are therefore used actively. With a forward and a backward pass being cached, therefore storing a state vector on both sides of each gate, the cache will have a size of 128~MiB for a 64-gate sample circuit in our 16-qubit case. Assuming parallelization over 32 cores, this results in slightly more than 4~GiB of total memory usage. In practice, about 25~\% more memory is required to account for the 16 vectors in the state vector array of the inner part of the Hessian computation.

\subsection{Time complexity}

The algorithm has exponential time complexity in the number of qubits $k$, as there are $2^k$ entries for each state vector and the summation for evaluating the trace in \cref{eq:f_target_sum_unit_vec} runs over $2^k$ unit vectors. Furthermore, we have to perform $n$ gate applications for a forward or backward pass, resulting in the following complexities:
\begin{align*}
\text{target function:} &\quad \mathcal{O}(n \, 2^k \, 2^k) = \mathcal{O}(n \, 4^k) \\
\text{gradient:} &\quad \mathcal{O}(n \, 2^k \, 2^k) = \mathcal{O}(n \, 4^k) \\
\text{Hessian:} &\quad \mathcal{O}(n^2 \, 2^k \, 2^k) = \mathcal{O}(n^2 \, 4^k)
\end{align*}
The dominant term as a function of the gates is the Hessian. In practice, we also observe that the Hessian computation takes up the biggest portion of the runtime, since the second forward pass involves a lot of intermediary state vectors.

\subsection{Parallelization}

The computational kernels described in \cref {sec:statevector-operations} are well-suited for vectorization to parallelize the inner loops. In addition, the ``embarrassingly'' parallel outer summation over the computational basis state vectors (see \cref{eq:f_target_sum_unit_vec}) via OpenMP multi-threading and MPI turned out to work well in our numerical experiments. This parallelization scheme also applies to the gradient and Hessian computation, since the calculations can be performed independently for each state vector.

On larger systems with a higher qubit count, the achieved speedup from parallelization is (slightly) below the increase in core count in experiments, see \cref{sec:parallelization_benchmark} below. This indicates that the main bottleneck is related to memory access instead of the arithmetic cost. The computational process reads and writes the entire state vector for each gate application and has to transfer this information from and to memory each time. For 16 qubits, a state vector requires 1~MiB, and thus, the gate application actively uses 2~MiB for the input and output vectors. A similar reasoning applies to the gate gradient computation, where two state vectors must be read from memory. Thus, the overall performance mainly depends on the number of state vectors that fit into the CPU's hardware caches. 

Using hardware with large L3 caches, such as server processors, is crucial for optimizing the method's performance. Hyper-threading can be used, as long as sufficient cache is available to store two state vectors for each thread. In modern server CPUs, it might be possible to fit both 1~MiB state vectors into L2 cache for even faster contraction.

\section{Evaluation}

This section contains benchmark evaluations of our implementation.

\subsection{Model system and circuit topology}

We first consider the spinless Fermi-Hubbard model on a one-dimensional lattice of size $L$ with periodic boundary conditions for our numerical experiments,
\begin{equation}
H_{\text{slFH}} = -J \sum_{j=0}^{L-1} \left( a^{\dagger}_j a_{j+1} + a^{\dagger}_{j+1} a_j \right) + U \sum_{j=0}^{L-1} n_{j} n_{j+1},
\end{equation}
where $a^{\dagger}_j$, $a_j$, $n_j = a^{\dagger}_j a_j$ are the fermionic creation, annihilation, and number operators acting on site $j$, and $J, U$ are real parameters. The first term in the Hamiltonian describes the kinetic hopping of fermions to neighboring lattice sites, and the second term describes the interaction between adjacent fermions. A Jordan-Wigner transformation maps the fermionic to bosonic operators:
\begin{subequations}
\begin{align}
a^{\dagger}_j &\to I_2 \otimes \cdots \otimes I_2 \otimes c^{\dagger} \otimes Z \otimes \cdots \otimes Z,\\
a_j &\to \underbrace{I_2 \otimes \cdots \otimes I_2}_{j \text{ terms}} \otimes c \otimes Z \otimes \cdots \otimes Z,
\end{align}
\end{subequations}
where $Z$ is the Pauli-$Z$ matrix, and $c^{\dagger}$ and $c$ are the local (bosonic) creation and annihilation operators:
\begin{equation}
c^{\dagger} = \begin{pmatrix} 0 & 0 \\ 1 & 0 \end{pmatrix}, \quad%
c = \begin{pmatrix} 0 & 1 \\ 0 & 0 \end{pmatrix},
\end{equation}
such that $c^{\dagger} \ket{0} = \ket{1}$ and $c \ket{1} = \ket{0}$.

A subtle issue arising from the Jordan-Wigner mapping is the breaking of translational invariance due to the kinetic term $(a^{\dagger}_{L-1} a_0 + a^{\dagger}_0 a_L)$, which introduces a chain of $Z$-strings running through the lattice. Since the brick wall circuit ansatz is translationally invariant by construction (with respect to shifts by two sites) and our method is intended to be applicable for large $L$, we bypass the issue by mapping $(a^{\dagger}_{L-1} a_0 + a^{\dagger}_0 a_L) \to (c^{\dagger}_{L-1} c_0 + c^{\dagger}_0 c_L)$, i.e., simply omitting the $Z$-strings for the periodic wrap-around. Following this procedure, the original Hamiltonian is transformed to its hard-core boson equivalent:
\begin{equation}
\label{eq:slFH_qubit}
\tilde{H}_{\text{slFH}} = -J \sum_{j=0}^{L-1} \left( c^{\dagger}_j c_{j+1} + c^{\dagger}_{j+1} c_j \right) + U \sum_{j=0}^{L-1} n_{j} n_{j+1}.
\end{equation}
Note the absence of any $Z$-operators.

As starting point for the circuit optimization, we use an even-odd Trotter splitting, $\tilde{H}_{\text{slFH}} = \tilde{H}_{\text{slFH}}^{\text{even}} + \tilde{H}_{\text{slFH}}^{\text{odd}}$, where the even and odd Hamiltonians contain the respective summands in \cref{eq:slFH_qubit} for even and odd $j$. The corresponding local two-site Hamiltonian has the matrix representation
\begin{equation}
h_{\text{loc}} = \begin{pmatrix} 0 & 0 & 0 & 0 \\ 0 & 0 & -J & 0 \\ 0 & -J & 0 & 0 \\ 0 & 0 & 0 & U \end{pmatrix}.
\end{equation}
The matrix exponential of $h_{\text{loc}}$ thus defines the two-qubit gates of a Trotter splitting approximation (which has a brick wall circuit layout):
\begin{equation}
\e^{-i t h_{\text{loc}}} = \begin{pmatrix} 1 & 0 & 0 & 0 \\ 0 & \cos(J t) & i \sin(J t) & 0 \\ 0 & i \sin(J t) & \cos(J t) & 0 \\ 0 & 0 & 0 & \e^{-i t U} \end{pmatrix}.
\end{equation}
As expected, it adheres to the sparsity structure of a parity-conserving gate as in \cref{eq:paritygatematrix}.

\subsection{Hardware aspects}

The optimization algorithm was tested on the LRZ CoolMuc-4 cluster. Each compute node has two  Intel(R) Xeon(R) Platinum 8480+ (Sapphire Rapids) processors within a dual socket setup, with a total of 112 cores and 210 MB of L3 cache. This allows for many state vectors to be cached, although when using a high number of threads, the L3 cache will still fill up quickly for circuits with 16 qubits.

\subsection{Parallelization benchmark}
\label{sec:parallelization_benchmark}

Our implementation in Ref.~\cite{matchgate_repo} supports parallelization (w.r.t.\ the basis vector summation) on HPC systems. On a single node, OpenMP is used to perform the optimization with up to 112 threads. To leverage more than one compute node, it is necessary to employ MPI communication protocols to synchronize the computations.

\begin{figure}[!ht]
\centering
\begin{subfigure}{0.45\textwidth} 
    \centering
    \includegraphics[width=\textwidth]{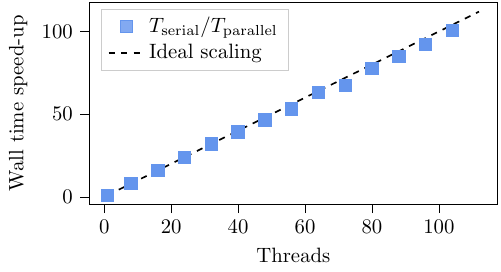} 
    \caption{OpenMP parallelization, $n_\text{qubits} = 10$, $n_\text{layers} = 11$}
    \label{fig:res:thread_scaling:sub2}
\end{subfigure}
\hspace{2pt}
\begin{subfigure}{0.45\textwidth}
    \centering
    \includegraphics[width=\textwidth]{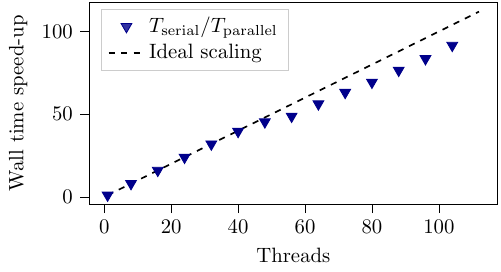} 
    \caption{OpenMP parallelization, $n_\text{qubits} = 12$, $n_\text{layers} = 3$}
    \label{fig:res:thread_scaling:sub1}
\end{subfigure}

\hspace{2pt}
\begin{subfigure}{0.45\textwidth} 
    \centering
    \includegraphics[width=\textwidth]{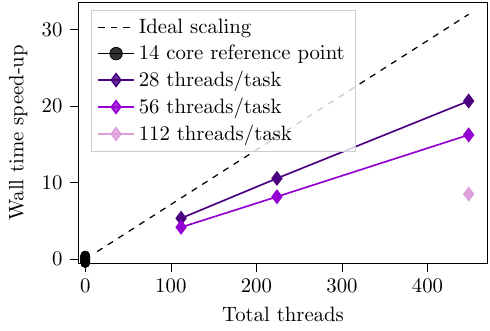} 
    \caption{OpenMP and MPI parallelization, $n_\text{qubits} = 16$, $n_\text{layers} = 11$.}
    \label{fig:res:thread_scaling:sub3}
\end{subfigure}

\caption{Multithreading benchmarks run on the LRZ CoolMuc-4 cluster. The dashed line represents the ideal linear scaling with slope one.}
\label{fig:res:thread_scaling}
\end{figure}

\cref{fig:res:thread_scaling} shows the speed-ups as a function of the number of threads allocated for different numbers of qubits and circuit layers. We observe a deviation from the ideal scaling when a certain number of threads is reached. This is likely due to cache misses, since the CPU cannot store all intermediate state vectors in the L3 memory. This effect is amplified for larger qubit numbers as each state vector requires more memory. Indeed, \cref{fig:res:thread_scaling:sub3} shows a benchmark for 16 qubits making use of MPI to utilize up to 448 cores, comparing the total execution time to the same optimization running on 14 threads, and the slope of the scaling is significantly below the ideal scaling. The scaling also depends on the ratio of MPI tasks to OpenMP threads, so it is advisable to perform separate benchmarks on a given platform to obtain the ideal configuration.

\subsection{Parity-conserving gates and invariance optimizations}

\begin{figure}[!ht]
\centering
\includegraphics[width=0.45\textwidth]{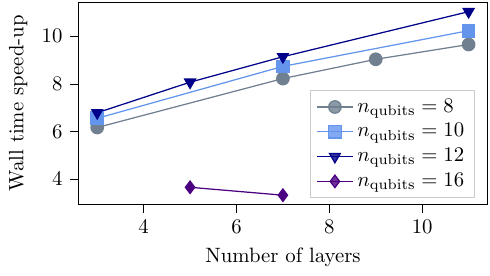}
\caption{Speed-up of the optimization procedure due to using parity-conserving gates.}
\label{fig:matchgate_speedups}
\end{figure}

Making use of the parity-conserving nature of the Hubbard model allows us to reduce the number of parameters to be optimized. Importantly, this reduction of degrees of freedom also reduces the number of temporary state vectors propagated in the Hessian computation. \cref{fig:matchgate_speedups} shows this speed-up when moving from general to sparse gates. When the L3 cache saturates for higher qubit numbers, this optimization is less significant as it does not reduce the number of actual data accesses.

\begin{figure}[!ht]
\centering
\includegraphics[width=0.45\textwidth]{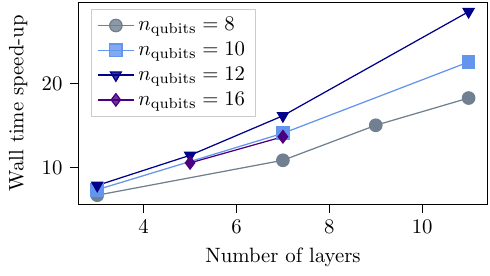}
\caption{Speed-up of the optimization procedure due to using parity-conserving gates and computing only translationally unique two-hole combinations.}
\label{fig:matchgate_invariance_speedups}
\end{figure}

Another optimization that can be applied to a brick wall circuit exploits the translational invariance of the system to reduce the number of gate hole pairs that need to be evaluated for the Hessian. If one gate hole pair can be transformed into another pair by a translation, then they result in the same numerical contribution to the Hessian block matrix. Thus, we run the computation for only one of these equivalent hole pairs and multiply by the number of appearances. \cref{fig:matchgate_invariance_speedups} shows the wall time speedup due to these optimizations. Interestingly, they also reduce the number of memory accesses to state vectors, thus not being as susceptible to L3 cache saturation. Note that the current implementation of this optimization requires a brick wall topology for the quantum circuit.

\subsection{Comparison with the \ac{MPO} method}
\label{sec:comparison}

The central tenet of the present work is a matrix-free formulation of Riemannian quantum circuit optimization to avoid storing the reference unitary $U$ as a dense matrix. Tensor network methods using \acp{MPO} to represent $U$ offer an alternative to this approach~\cite{Le2025riemannian}. Here, we compare these two approaches in terms of computational time. We have implemented a first-order Riemannian adaptation of the ADAM optimizer in the \ac{MPO} version~\cite{kingma2014adam, Le2025riemannian}, and thus we focus on the gradient evaluation time for our comparison. The state vector approach is expected to run faster for small system sizes due to the relatively simple arithmetic operations -- for example, without the need of singular value decompositions as required by the \ac{MPO} method -- and conversely, one expects an advantage of the \ac{MPO} approach for larger systems since it avoids the curse of dimensionality.

For this comparison, we consider the one-dimensional spinful Fermi-Hubbard model:
\begin{equation}\label{eq:sfFH}
H_\text{sfFH} = -J \sum_{\langle j, \ell \rangle, \sigma} \left( a_{j \sigma}^\dagger a_{\ell \sigma} + a_{\ell \sigma}^\dagger a_{j \sigma} \right) + U \sum_j n_{j \uparrow} n_{j \downarrow},
\end{equation}
with $J = 1$, $U = 4$, $\sigma \in \{\uparrow, \downarrow\}$, denoting the spin, and periodic boundary conditions. When enumerating all spin-up sites first, followed by the spin-down sites, we can avoid any $Z$-operators after the Jordan-Wigner transformation, analogous to the spinless model. Note that the interaction term of the Hamiltonian results in long-range two-qubit gates for this enumeration.

We have implemented periodic boundary conditions in our \ac{MPO} method via a long-range interaction from the first to the last site. In this analysis, we cap the maximum virtual bond dimension at $\chi_\text{max} = 256$.

This comparison is qualitative rather than quantitative, as the implementation of the current \ac{MPO} is not optimized for HPC systems. 

\subsubsection{Scaling of gradient evaluation time}

\begin{figure}
\centering
\includegraphics[width=0.49\textwidth]{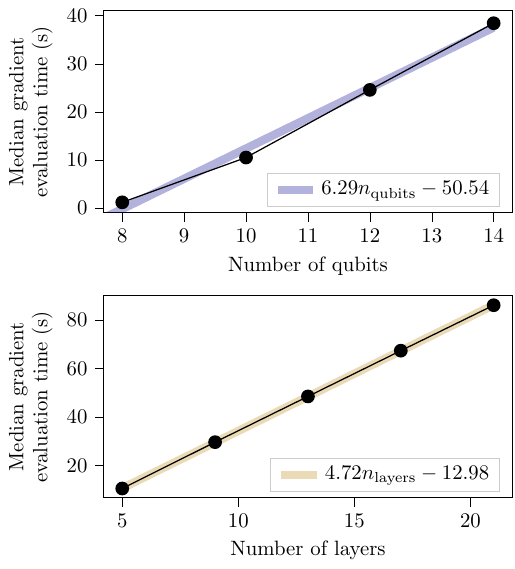}
\caption{Gradient evaluation time using the \ac{MPO} method. The median evaluation time for 100 random quantum gate initializations is shown for various numbers of qubits $n_\text{qubits}$ and layers $n_\text{layers}$, exhibiting a linear scaling for each.}
\label{fig:mpo-benchmark-time}
\end{figure}

\begin{figure}
\centering
\includegraphics[width=0.49\textwidth]{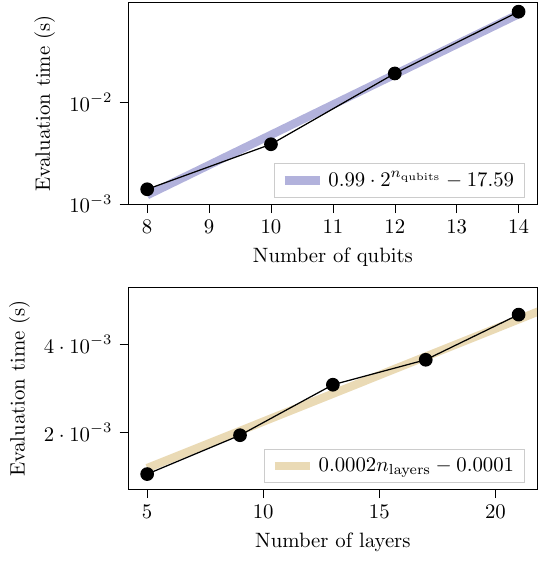}
\caption{Evaluation time for one summand of the gradient computation depending on the number of qubits and circuit layers for the state vector method, exhibiting an exponential scaling with respect to $n_\text{qubits}$. (Note the logarithmic scale in the top subfigure.)}
\label{fig:matchgate_gradient}
\end{figure}

In the following, we analyze the evaluation time of gradient computations for various numbers of qubits, $n_\text{qubits}$, and layers, $n_\text{layers}$. 100 random quantum gate initializations are generated for each data point, and the median evaluation time is presented. The results for the \ac{MPO} method are shown in \cref{fig:mpo-benchmark-time}. The timings for the state vector algorithm and a single state vector summand (see~\cref{eq:f_target_sum_unit_vec}) are presented in \cref{fig:matchgate_gradient}. The numbers in \cref{fig:matchgate_gradient} thus need to be scaled by $2^{n_\text{qubits}}$ to obtain the overall gradient evaluation time.

In order to compare the gradient evaluation scaling for an increasing number of qubits, we chose corresponding brick wall circuits with 5 layers, respectively. 
For the comparison with respect to an increasing number of layers, the system size was set to 10 qubits. 

As shown, the gradient computation within the \ac{MPO} method scales linearly with $n_\text{qubits}$ and $n_\text{layers}$, whereas the state vector approach scales exponentially with the number of qubits, as expected. Consequently, the \ac{MPO} method is a suitable alternative for optimizing larger systems directly, especially when missing translational invariance.

\subsubsection{Convergence comparison}

Since the trust-region algorithm is a second-order optimization routine, it is expected to converge faster than the first-order Riemannian ADAM algorithm. Additionally, utilizing the Hessian reduces the risk of encountering local minima. To verify this, we compare the convergence behavior of both methods for the spinful Fermi-Hubbard model given in \cref{eq:sfFH} on 8 sites.

\begin{figure}
    \centering
    \includegraphics[width=0.95\linewidth]{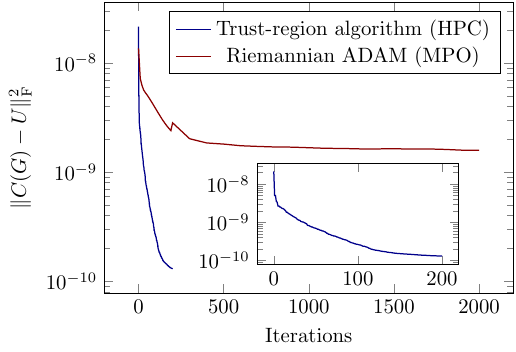}
    \caption{Comparison of the convergence behavior of the second-order trust-region algorithm and the first-order Riemannian ADAM. A brick wall circuit with 21 layers was optimized to approximate the time evolution operator of the spinful Fermi-Hubbard model with periodic boundary conditions on 8 sites.}
    \label{fig:convergence-benchmark}
\end{figure}

The quantum circuit is initialized as a fourth-order Trotter step, resulting in 21 layers. Due to practical reasons, the implementations differ in orbital ordering, leading to slightly different initializations. The results presented in \cref{fig:convergence-benchmark} demonstrate a significant convergence advantage for the second-order method. 
It requires only 200 update steps to reduce the approximation error by two orders of magnitude. Namely, an initial error of $\epsilon_\text{init} = 2.15\cdot 10^{-8}$ was reduced to $\epsilon_\text{opt} = 1.30\cdot 10^{-10}$. 
In contrast, the Riemannian ADAM method took 2000 optimization steps to converge and did not find the same minimum, reducing the approximation error by only one order of magnitude. 
Namely, an initial error of $\epsilon_\text{init} = 1.36\cdot 10^{-8}$ was reduced to $\epsilon_\text{opt} = 1.57\cdot 10^{-9}$. 

In conclusion, while our HPC implementation of the trust-region algorithm is not scalable for non-translationally invariant systems, it offers a fast-converging and robust optimization method for simulating small systems. 
In the future, we want to investigate how to promote the \ac{MPO} method to a second-order Riemannian optimization.

\section{Conclusions and outlook}

We have formulated the detailed computational steps for performing state vector-based, matrix-free Riemannian quantum circuit optimization using the Hessian and the trust-region algorithm. Our main technical contributions are the computational kernels for the algorithmic operations in \cref{sec:statevector-operations}.

We have demonstrated a significant performance gain of more than an order of magnitude from exploiting the parity-conserving gate sparsity structure and translational invariance. While the sparsity pattern naturally emerges from a Trotter splitting, lifting the sparsity restriction could, in principle, lead to more expressive circuits and better approximation accuracies. Interestingly, in our preparatory numerical experiments, we have found that this is not the case (for parity-conserving Hamiltonians), i.e., the sparse gates achieve the same accuracy. A theoretical investigation could be a promising endeavor for future work.

A generalization to two-dimensional lattice topologies constitutes another relevant project for future work. Our state vector approach can realistically work on a $4 \times 4$ lattice, for example. Note, however, that exploiting translation invariance in two dimensions leads to less cost savings as compared to the one-dimensional case: for example, on a $4 \times 4$ lattice, one can only translate by two sites in one (or both) coordinate directions, whereas on a one-dimensional lattice with $16$ sites, the redundant translations are $2, 4, \dots, 14$ sites.

Regarding parallelization by multithreading, we have used the ``embarrassingly parallel'' computation of the sum over basis states in \cref{eq:f_target_sum_unit_vec}. Alternatively, parallelizing the loops of the computational kernels in \cref{sec:statevector-operations} is conceivable as well and could lead to a more sequential memory access, but is unlikely to improve cache performance, as the access occurs within four sequential streams. Our observations showed worse performance compared to the sum over basis states variant. This is likely due to task management / distribution as well as cache lines getting evicted and fetched as they are being accessed by different cores.

Related to performance optimization, a natural extension of our work is an implementation targeted at hardware accelerators such as graphics processing units (GPUs) or tensor processing units (TPUs).

Our qualitative computational performance comparison in \cref{sec:comparison} with an alternative implementation using an \ac{MPO} representation of the target unitary time evolution operator leads to the following insight: our state vector-based method is considerably faster for small system sizes, but will eventually be overtaken by the \ac{MPO} approach for larger systems due to the curse of dimensionality. The \ac{MPO} approach is immune to the curse of dimensionality if an area law of the entanglement entropy holds, which is expected to be the case for one-dimensional lattices, a local Hamiltonian, and a fixed evolution time. Consequently, there will be a break-even point at a certain system size regarding the computational performance, and whether to preferentially use the statevector- or \ac{MPO}-based algorithm will depend on whether this system size is sufficiently large for the considered setup (i.e., whether spurious wrap-around influences are avoided, see the light cone picture in Ref.~\cite{Kotil2024}). Note that a quantitative performance comparison also depends on implementation and code optimization details, and that we have only compared the gradient evaluation step here.

Evaluating second derivatives to obtain the Hessian matrix and using it for the Riemannian trust-region algorithm can significantly reduce the number of iterations and reach a better approximation accuracy, as demonstrated in \cref{fig:convergence-benchmark}. However, directly computing the Hessian matrix via the \ac{MPO} framework is rather intricate. A promising solution is to evaluate Hessian-vector products instead. We outline the computational steps in \cref{sec:Hessian-vector_product}.

In a follow-up work, we plan to approximate the sum in \cref{eq:f_target_sum_unit_vec} by sampling over random vectors. This aligns very well with tensor network methods when representing these states as \acp{MPS}, combined with Hessian-vector products.

\subsection*{Data availability}

The source code for our parallelized C-implementation exploiting the parity-conserving gate structure and translation invariance is available at~\cite{matchgate_repo}. It is based on the preliminary code base (without parallelization) at~\cite{rqcopt_hpc}. An alternative C++ implementation for brick wall topologies is accessible at~\cite{rqcopt_cpp}. A reference (non-optimized) Python version developed for~\cite{Kotil2024} can be found at~\cite{rqcopt}. 
The implementation of the alternative first-order \ac{MPO} method is published at~\cite{rqcopt_mpo}.

\acknowledgments

We would like to thank Roeland Wiersema for insightful discussions.

We thank the Leibniz Supercomputing Centre (LRZ) for providing computing resources and the Munich Center for Quantum Science and Technology (MCQST). Q.H. is supported by the Bavarian Ministry of Economic Affairs, Regional Development and Energy via the project BayQS with funds from the Hightech Agenda Bayern.

\appendix

\section{Evaluating Hessian-vector products}
\label{sec:Hessian-vector_product}

This section describes an algorithm for computing circuit Hessian-vector products at a cost scaling linearly with the number of gates. 
Since the inner truncated-CG iteration of the Riemannian optimization would require many such Hessian-vector product evaluations, this approach turned out to be computationally slower than calculating the Hessian matrix once (as discussed in the main text). 
Nevertheless, it might be of independent interest and prove advantageous in a different setting.

\begin{figure}[!ht]
\centering
\input{gfx/directed_gradient_graph.tikz}
\caption{Computational graph for evaluating the directed gradient (with respect to the circuit gates), where the direction is stored in the matrices $Z_{\ell}$.}
\label{fig:directed_gradient_graph}
\end{figure}
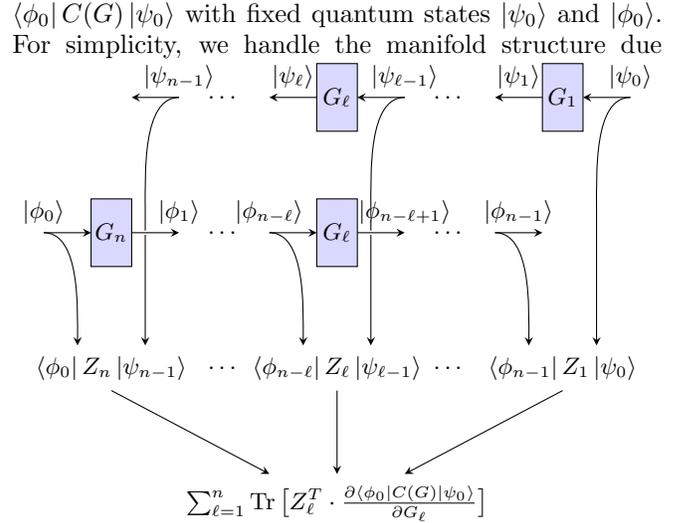

In general, denoting the cost function by $f$, the goal is to evaluate $H_f \cdot z$, where $H_f$ is the Hessian matrix of second derivatives of $f$, and $z$ is a specified (arbitrary) direction vector of the same dimension as the input to $f$. As noted in the automatic differentiation (AD) research community \cite{Bishop1992, Christianson1992, GriewankWalther2008}, $H_f \cdot z$ can be computed without explicitly forming the Hessian matrix at the same asymptotic cost (up to constant prefactors) as a forward pass through the neural network. Briefly, the idea consists in first building the computational graph for the derivative of $f$ in direction $z$ via forward-mode AD, regarded as a new function: $g(x) \coloneqq D f(x) \cdot z$. Next, one obtains the derivative of $g$ via backpropagation (with the same computational complexity as evaluating $f$), which equals the sought Hessian-vector product: $D g = H_f \cdot z$.

We now realize this prescription for the present case of quantum circuits. As before, we start with one summand of the overall target function in \cref{eq:f_target_sum_unit_vec} as our gate-dependent target function, which we write as $\bra{\phi_0} C(G) \ket{\psi_0}$ with fixed quantum states $\ket{\psi_0}$ and $\ket{\phi_0}$. For simplicity, we handle the manifold structure due to the unitary property of the quantum gates $\{ G_{\ell} \}$ later and regard the gates as general complex matrices, $G_{\ell} \in \C^{m \times m}$. We also work with a complex-valued target function here (which is actually holomorphic as a function of the gate parameters) and take complex derivatives.

Our goal is to evaluate the Hessian-vector product $H \cdot \mathrm{vec}(Z)$, where the Hessian $H = \frac{\partial^2 \bra{\phi_0} C(G) \ket{\psi_0}}{\partial G_{\ell} \partial G_{\ell'}}$ collects all second derivatives with respect to the gate entries, and $Z = (Z_1, \dots, Z_n)$ with $Z_{\ell} \in \C^{m \times m}$ is the (arbitrary) derivative direction for gate $G_{\ell}$.

As the first step, we identify the computational graph (shown in \cref{fig:directed_gradient_graph}) for evaluating the directed gradient
\begin{equation}\sum_{\ell=1}^n \Tr\bigg[Z_{\ell}^T \cdot \frac{\partial \bra{\phi_0} C(G) \ket{\psi_0}}{\partial G_{\ell}}\bigg],
\end{equation}
which is a function mapping the gates to a complex number and corresponds to the $g$ function from above. 
With our conventions,
\begin{equation}
\Tr\bigg[Z_{\ell}^T \cdot \frac{\partial \bra{\phi_0} C(G) \ket{\psi_0}}{\partial G_{\ell}}\bigg] = \bra{\phi_{n-\ell}} Z_{\ell} \ket{\psi_{\ell-1}}.
\end{equation}
As the second step, one derives this function once more with respect to the gate entries, yielding $H \cdot \mathrm{vec}(Z)$. In practice, this derivative can be computed by a backward pass through the computational graph, i.e., following the directed arrows in \cref{fig:directed_gradient_graph} in reverse direction. Note that each arrow branching becomes a summation in this backward pass.

\bibliography{references}

\end{document}

%% file: gfx/generic_quantum_circuit.tikz
\begin{tikzpicture}[scale=0.6, >=stealth]
\foreach \y in {0, ..., 4}
{
    \draw (-1.5*0.5, \y) -- ( 1.5*6.5, \y);
}
\draw[fill=gatecolor] ( 1.5*6-\gw, 3.5-\gh) rectangle ( 1.5*6+\gw, 3.5+\gh);
\node at ( 1.5*6, 3.5) {$G_1$};
\draw[fill=gatecolor] ( 1.5*5-\gw, 2.5) -- ( 1.5*5-\gw, 2.5+\gh) -- ( 1.5*5+\gw, 2.5+\gh) -- ( 1.5*5+\gw, 2.5);
\draw[fill=gatecolor] ( 1.5*5+\gw, 1.5) -- ( 1.5*5+\gw, 1.5-\gh) -- ( 1.5*5-\gw, 1.5-\gh) -- ( 1.5*5-\gw, 1.5);
\node at ( 1.5*5, 1.5) {$G_2$};
\draw[fill=gatecolor] ( 1.5*4-\gw, 2.5-\gh) rectangle ( 1.5*4+\gw, 2.5+\gh);
\node at ( 1.5*4, 2.5) {$G_3$};
\draw[fill=gatecolor] ( 1.5*3-\gw,-0.5) -- ( 1.5*3-\gw,-0.5+\gh) -- ( 1.5*3+\gw,-0.5+\gh) -- ( 1.5*3+\gw,-0.5);
\draw[fill=gatecolor] ( 1.5*3+\gw, 4.5) -- ( 1.5*3+\gw, 4.5-\gh) -- ( 1.5*3-\gw, 4.5-\gh) -- ( 1.5*3-\gw, 4.5);
\node at ( 1.5*3, 4.5) {$G_4$};
\draw[fill=gatecolor] ( 1.5*2-\gw, 1.5-\gh) rectangle ( 1.5*2+\gw, 1.5+\gh);
\node at ( 1.5*2, 1.5) {$G_5$};
\draw[fill=gatecolor] ( 1.5*1-\gw, 0.5-\gh) rectangle ( 1.5*1+\gw, 0.5+\gh);
\node at ( 1.5*1, 0.5) {$G_6$};
\draw[fill=gatecolor] ( 1.5*0-\gw, 3.5-\gh) rectangle ( 1.5*0+\gw, 3.5+\gh);
\node at ( 1.5*0, 3.5) {$G_7$};
\draw[decorate, decoration={brace,mirror,amplitude=4}, thick] (-0.75,-1) -- ( 1.5*6+0.75,-1);
\node at ( 1.5*3,-2) {$C(G_1, \dots, G_7)$};
\draw[decorate, decoration={brace,mirror,amplitude=4}, thick] ( 1.5*6.75, -0.25) -- ( 1.5*6.75, 4.25);
\node at ( 1.5*7.25, 2) {$\ket{\psi}$};
\end{tikzpicture}

%% file: gfx/target_network.tikz
\begin{tikzpicture}[scale=0.6, >=stealth]
\foreach \y in {0, ..., 4}
{
    \draw (-1.5*1+\gw, \y) -- ( 1.5*7-\gw, \y);
}
\draw[fill=gatecolor] ( 1.5*6-\gw, 3.5-\gh) rectangle ( 1.5*6+\gw, 3.5+\gh);
\draw[fill=gatecolor] ( 1.5*5-\gw, 2.5) -- ( 1.5*5-\gw, 2.5+\gh) -- ( 1.5*5+\gw, 2.5+\gh) -- ( 1.5*5+\gw, 2.5);
\draw[fill=gatecolor] ( 1.5*5+\gw, 1.5) -- ( 1.5*5+\gw, 1.5-\gh) -- ( 1.5*5-\gw, 1.5-\gh) -- ( 1.5*5-\gw, 1.5);
\draw[fill=gatecolor] ( 1.5*4-\gw, 2.5-\gh) rectangle ( 1.5*4+\gw, 2.5+\gh);
\draw[fill=gatecolor] ( 1.5*3-\gw,-0.5) -- ( 1.5*3-\gw,-0.5+\gh) -- ( 1.5*3+\gw,-0.5+\gh) -- ( 1.5*3+\gw,-0.5);
\draw[fill=gatecolor] ( 1.5*3+\gw, 4.5) -- ( 1.5*3+\gw, 4.5-\gh) -- ( 1.5*3-\gw, 4.5-\gh) -- ( 1.5*3-\gw, 4.5);
\draw[fill=gatecolor] ( 1.5*2-\gw, 1.5-\gh) rectangle ( 1.5*2+\gw, 1.5+\gh);
\draw[fill=gatecolor] ( 1.5*1-\gw, 0.5-\gh) rectangle ( 1.5*1+\gw, 0.5+\gh);
\draw[fill=gatecolor] ( 1.5*0-\gw, 3.5-\gh) rectangle ( 1.5*0+\gw, 3.5+\gh);
\draw[fill=statecolor, rounded corners] (1.5*7-\gw, -0.5) rectangle ( 1.5*7+\gw, 4.5);
\node at (1.5*7, 2) {$\ket{\psi}$};
\draw[fill=statecolor, rounded corners] (-1.5*1-\gw, -0.5) rectangle (-1.5*1+\gw, 4.5);
\node at (-1.5*1, 2) {$\ket{\phi}$};
\end{tikzpicture}

%% file: gfx/network_gradient_hole.tikz
\begin{tikzpicture}[scale=0.6, >=stealth]
\foreach \y in {0, ..., 4}
{
    \draw (-1.5*1+\gw, \y) -- ( 1.5*7-\gw, \y);
}
\draw[fill=gatecolor] ( 1.5*6-\gw, 3.5-\gh) rectangle ( 1.5*6+\gw, 3.5+\gh);
\draw[dotted, fill=white] ( 1.5*5-\gw, 2.5) -- ( 1.5*5-\gw, 2.5+\gh) -- ( 1.5*5+\gw, 2.5+\gh) -- ( 1.5*5+\gw, 2.5);
\draw[dotted, fill=white] ( 1.5*5+\gw, 1.5) -- ( 1.5*5+\gw, 1.5-\gh) -- ( 1.5*5-\gw, 1.5-\gh) -- ( 1.5*5-\gw, 1.5);
\node at ( 1.5*5, 1.5) {$G_{\ell}$};
\draw[fill=gatecolor] ( 1.5*4-\gw, 2.5-\gh) rectangle ( 1.5*4+\gw, 2.5+\gh);
\draw[fill=gatecolor] ( 1.5*3-\gw,-0.5) -- ( 1.5*3-\gw,-0.5+\gh) -- ( 1.5*3+\gw,-0.5+\gh) -- ( 1.5*3+\gw,-0.5);
\draw[fill=gatecolor] ( 1.5*3+\gw, 4.5) -- ( 1.5*3+\gw, 4.5-\gh) -- ( 1.5*3-\gw, 4.5-\gh) -- ( 1.5*3-\gw, 4.5);
\draw[fill=gatecolor] ( 1.5*2-\gw, 1.5-\gh) rectangle ( 1.5*2+\gw, 1.5+\gh);
\draw[fill=gatecolor] ( 1.5*1-\gw, 0.5-\gh) rectangle ( 1.5*1+\gw, 0.5+\gh);
\draw[fill=gatecolor] ( 1.5*0-\gw, 3.5-\gh) rectangle ( 1.5*0+\gw, 3.5+\gh);
\draw[fill=statecolor, rounded corners] (1.5*7-\gw, -0.5) rectangle ( 1.5*7+\gw, 4.5);
\node at (1.5*7, 2) {$\ket{\psi}$};
\draw[fill=statecolor, rounded corners] (-1.5*1-\gw, -0.5) rectangle (-1.5*1+\gw, 4.5);
\node at (-1.5*1, 2) {$\ket{\phi}$};
\end{tikzpicture}

%% file: gfx/network_hessian_holes.tikz
\begin{tikzpicture}[scale=0.6, >=stealth]
\foreach \y in {0, ..., 4}
{
    \draw (-1.5*1+\gw, \y) -- ( 1.5*7-\gw, \y);
}
\draw[fill=gatecolor] ( 1.5*6-\gw, 3.5-\gh) rectangle ( 1.5*6+\gw, 3.5+\gh);
\draw[dotted, fill=white] ( 1.5*5-\gw, 2.5) -- ( 1.5*5-\gw, 2.5+\gh) -- ( 1.5*5+\gw, 2.5+\gh) -- ( 1.5*5+\gw, 2.5);
\draw[dotted, fill=white] ( 1.5*5+\gw, 1.5) -- ( 1.5*5+\gw, 1.5-\gh) -- ( 1.5*5-\gw, 1.5-\gh) -- ( 1.5*5-\gw, 1.5);
\node at ( 1.5*5, 1.5) {$G_{\ell}$};
\draw[fill=gatecolor] ( 1.5*4-\gw, 2.5-\gh) rectangle ( 1.5*4+\gw, 2.5+\gh);
\draw[fill=gatecolor] ( 1.5*3-\gw,-0.5) -- ( 1.5*3-\gw,-0.5+\gh) -- ( 1.5*3+\gw,-0.5+\gh) -- ( 1.5*3+\gw,-0.5);
\draw[fill=gatecolor] ( 1.5*3+\gw, 4.5) -- ( 1.5*3+\gw, 4.5-\gh) -- ( 1.5*3-\gw, 4.5-\gh) -- ( 1.5*3-\gw, 4.5);
\draw[dotted, fill=white] ( 1.5*2-\gw, 1.5-\gh) rectangle ( 1.5*2+\gw, 1.5+\gh);
\node at ( 1.5*2, 1.5) {$G_{\ell'}$};
\draw[fill=gatecolor] ( 1.5*1-\gw, 0.5-\gh) rectangle ( 1.5*1+\gw, 0.5+\gh);
\draw[fill=gatecolor] ( 1.5*0-\gw, 3.5-\gh) rectangle ( 1.5*0+\gw, 3.5+\gh);
\draw[fill=statecolor, rounded corners] (1.5*7-\gw, -0.5) rectangle ( 1.5*7+\gw, 4.5);
\node at (1.5*7, 2) {$\ket{\psi}$};
\draw[fill=statecolor, rounded corners] (-1.5*1-\gw, -0.5) rectangle (-1.5*1+\gw, 4.5);
\node at (-1.5*1, 2) {$\ket{\phi}$};
\end{tikzpicture}

%% file: gfx/forward_pass.tikz
\begin{tikzpicture}[scale=0.6, >=stealth]
\foreach \y in {0, ..., 4}
{
    \draw ( 1.5*2+\gw, \y) -- ( 1.5*7-\gw, \y);
}
\node at ( 1.5*1.5, 2) {$\dots$};
\draw[fill=gatecolor] ( 1.5*6-\gw, 3.5-\gh) rectangle ( 1.5*6+\gw, 3.5+\gh);
\draw[fill=gatecolor] ( 1.5*5-\gw, 2.5) -- ( 1.5*5-\gw, 2.5+\gh) -- ( 1.5*5+\gw, 2.5+\gh) -- ( 1.5*5+\gw, 2.5);
\draw[fill=gatecolor] ( 1.5*5+\gw, 1.5) -- ( 1.5*5+\gw, 1.5-\gh) -- ( 1.5*5-\gw, 1.5-\gh) -- ( 1.5*5-\gw, 1.5);
\draw[fill=gatecolor] ( 1.5*4-\gw, 2.5-\gh) rectangle ( 1.5*4+\gw, 2.5+\gh);
\draw[fill=gatecolor] ( 1.5*3-\gw,-0.5) -- ( 1.5*3-\gw,-0.5+\gh) -- ( 1.5*3+\gw,-0.5+\gh) -- ( 1.5*3+\gw,-0.5);
\draw[fill=gatecolor] ( 1.5*3+\gw, 4.5) -- ( 1.5*3+\gw, 4.5-\gh) -- ( 1.5*3-\gw, 4.5-\gh) -- ( 1.5*3-\gw, 4.5);
\draw[fill=statecolor, rounded corners] (1.5*7-\gw, -0.5) rectangle ( 1.5*7+\gw, 4.5);
\node at (1.5*7, 2) {$\ket{\psi}$};
\foreach \i in {1, ..., 4}
{
    \pgfmathsetmacro{\x}{6.5-\i};
    \draw[dashed] (1.5*\x, 4.5) -- (1.5*\x, -0.5) node[below] {$\ket{\psi_\i}$};
}
\draw[->, thick] ( 1.5*7-\gw, 5.5) -- node[below] {forward} ( 1.5*1+\gw, 5.5);
\end{tikzpicture}

%% file: gfx/backward_pass.tikz
\begin{tikzpicture}[scale=0.6, >=stealth]
\foreach \y in {0, ..., 4}
{
    \draw (-1.5*1+\gw, \y) -- ( 1.5*2-\gw, \y);
    \draw ( 1.5*3+\gw, \y) -- ( 1.5*5-\gw, \y);
}
\node at ( 1.5*2.5, 2) {$\dots$};
\node at ( 1.5*5.5, 2) {$\dots$};
\draw[dotted, fill=white] ( 1.5*4-\gw, 2.5) -- ( 1.5*4-\gw, 2.5+\gh) -- ( 1.5*4+\gw, 2.5+\gh) -- ( 1.5*4+\gw, 2.5);
\draw[dotted, fill=white] ( 1.5*4+\gw, 1.5) -- ( 1.5*4+\gw, 1.5-\gh) -- ( 1.5*4-\gw, 1.5-\gh) -- ( 1.5*4-\gw, 1.5);
\node at ( 1.5*4, 1.5) {$G_{\ell}$};
\draw[fill=gatecolor] ( 1.5*1-\gw, 0.5-\gh) rectangle ( 1.5*1+\gw, 0.5+\gh);
\draw[fill=gatecolor] ( 1.5*0-\gw, 3.5-\gh) rectangle ( 1.5*0+\gw, 3.5+\gh);
\draw[fill=statecolor, rounded corners] (-1.5*1-\gw, -0.5) rectangle (-1.5*1+\gw, 4.5);
\node at (-1.5*1, 2) {$\ket{\phi}$};
\foreach \i in {1, ..., 2}
{
    \pgfmathsetmacro{\x}{-0.5+\i};
    \draw[dashed] (1.5*\x, 4.5) -- (1.5*\x, -0.5) node[below] {$\ket{\phi_\i}$};
}
\draw[dashed] (1.5*3.5, 4.5) -- (1.5*3.5, -0.5) node[below] {$\ket{\phi_{n-\ell}}$};
\draw[dashed] (1.5*4.5, 4.5) -- (1.5*4.5, -0.5) node[below] {$\ket{\psi_{\ell-1}}$};
\draw[->, thick] (-1.5*1+\gw, 5.5) -- node[below] {backward} ( 1.5*5-\gw, 5.5);

\draw[decorate, decoration={brace,mirror,amplitude=4}, thick] ( 1.5*3, -2) -- ( 1.5*5, -2);

\begin{scope}[shift={( 0, -7.5)}]
\foreach \y in {0, ..., 4}
{
    \draw ( 1.5*3+\gw, \y) -- ( 1.5*5-\gw, \y);
}
\node at ( 1.5*0.5, 2) {$\partial\bra{\phi} C(G) \ket{\psi} / \partial G_{\ell} =$};
\draw[dotted, fill=white] ( 1.5*4-\gw, 2.5) -- ( 1.5*4-\gw, 2.5+\gh) -- ( 1.5*4+\gw, 2.5+\gh) -- ( 1.5*4+\gw, 2.5);
\draw[dotted, fill=white] ( 1.5*4+\gw, 1.5) -- ( 1.5*4+\gw, 1.5-\gh) -- ( 1.5*4-\gw, 1.5-\gh) -- ( 1.5*4-\gw, 1.5);
\draw[fill=statecolor, rounded corners] (1.5*5-\gw, -0.5) rectangle ( 1.5*5+\gw, 4.5);
\node[rotate=90] at ( 1.5*5, 2) {$\ket{\psi_{\ell-1}}$};
\draw[fill=statecolor, rounded corners] ( 1.5*3-\gw, -0.5) rectangle ( 1.5*3+\gw, 4.5);
\node[rotate=90] at ( 1.5*3, 2) {$\ket{\phi_{n-\ell}}$};
\end{scope}
\end{tikzpicture}

%% file: gfx/hole_application.tikz
\begin{tikzpicture}[scale=0.6, >=stealth]
\foreach \y in {0, ..., 4}
{
    \draw (-1.5*2+\gw, \y) -- (-\gw, \y);
}
\draw[dotted, fill=white] (-1.5-\gw, 2.5) -- (-1.5-\gw, 2.5+\gh) -- (-1.5+\gw, 2.5+\gh) -- (-1.5+\gw, 2.5);
\draw[dotted, fill=white] (-1.5-\gw, 1.5) -- (-1.5-\gw, 1.5-\gh) -- (-1.5+\gw, 1.5-\gh) -- (-1.5+\gw, 1.5);
\draw[mblue,   very thick] (-1.5*1.5, 3) -- (-1.5-\gw, 3);
\draw[morange, very thick] (-1.5*1.5, 1) -- (-1.5-\gw, 1);
\draw[mgreen,  very thick] (-\gw,     3) -- (-1.5+\gw, 3);
\draw[mred,    very thick] (-\gw,     1) -- (-1.5+\gw, 1);
\draw[fill=statecolor, rounded corners] (-\gw, -0.5) rectangle ( \gw, 4.5);
\node[rotate=90] at (0, 2) {$\ket{\psi_{\ell-1}}$};
\node at (2.25, 2) {$=$};
\begin{scope}[shift={(5,0)}]
\foreach \y in {0, ..., 4}
{
    \draw (-1.5+\gw, \y) -- (-\gw, \y);
}
\draw[fill=statecolor, rounded corners] (-\gw, -0.5) rectangle ( \gw, 4.5);
\node[rotate=90] at (0, 2) {$\ket{\psi^{(\ell)}_{\ell}}$};
\draw[mblue,   very thick] ( \gw, 2.3) -- ( 1.5-\gw, 2.3);
\draw[morange, very thick] ( \gw, 2.1) -- ( 1.5-\gw, 2.1);
\draw[mgreen,  very thick] ( \gw, 1.9) -- ( 1.5-\gw, 1.9);
\draw[mred,    very thick] ( \gw, 1.7) -- ( 1.5-\gw, 1.7);
\end{scope}
\end{tikzpicture}

%% file: gfx/hessian_pass.tikz
\begin{tikzpicture}[scale=0.6, >=stealth]
\foreach \y in {0, ..., 4}
{
    \draw ( 1.5*2+\gw, \y) -- ( 1.5*5-\gw, \y);
    \draw (-1.5*1+\gw, \y) -- ( 1.5*1-\gw, \y);
}
\node at ( 1.5*1.5, 2) {$\dots$};
\draw[fill=gatecolor] ( 1.5*4-\gw, 2.5-\gh) rectangle ( 1.5*4+\gw, 2.5+\gh);
\draw[fill=gatecolor] ( 1.5*3-\gw,-0.5) -- ( 1.5*3-\gw,-0.5+\gh) -- ( 1.5*3+\gw,-0.5+\gh) -- ( 1.5*3+\gw,-0.5);
\draw[fill=gatecolor] ( 1.5*3+\gw, 4.5) -- ( 1.5*3+\gw, 4.5-\gh) -- ( 1.5*3-\gw, 4.5-\gh) -- ( 1.5*3-\gw, 4.5);
\draw[dotted, fill=white] ( 1.5*0-\gw, 1.5-\gh) rectangle ( 1.5*0+\gw, 1.5+\gh);
\node at ( 1.5*0, 1.5) {$G_{\ell'}$};
\draw[fill=statecolor, rounded corners] (1.5*5-\gw, -0.5) rectangle ( 1.5*5+\gw, 4.5);
\node[rotate=90] at (1.5*5, 2) {$\ket{\psi^{(\ell)}_{\ell}}$};
\draw ( 1.5*5+\gw, 2.3) -- ( 1.5*6-\gw, 2.3);
\draw ( 1.5*5+\gw, 2.1) -- ( 1.5*6-\gw, 2.1);
\draw ( 1.5*5+\gw, 1.9) -- ( 1.5*6-\gw, 1.9);
\draw ( 1.5*5+\gw, 1.7) -- ( 1.5*6-\gw, 1.7);
\draw[dashed] ( 1.5*3.5, 4.5) -- ( 1.5*3.5, -0.5) node[below] {$\ket{\psi^{(\ell)}_{\ell+1}}$};
\draw[dashed] ( 1.5*2.5, 4.5) -- ( 1.5*2.5, -0.5) node[below] {$\ket{\psi^{(\ell)}_{\ell+2}}$};
\draw[dashed] ( 1.5*0.5, 4.5) -- ( 1.5*0.5, -0.5) node[below] {\quad $\ket{\psi^{(\ell)}_{\ell'-1}}$};
\draw[dashed] (-1.5*0.5, 4.5) -- (-1.5*0.5, -0.5) node[below] {$\ket{\phi_{n-\ell'}}$};
\draw[->, thick] ( 1.5*5-\gw, 5.5) -- node[below] {second forward pass} (-1.5*1+\gw, 5.5);

\draw[decorate, decoration={brace,mirror,amplitude=4}, thick] (-1.5*1, -2) -- ( 1.5*1, -2);

\begin{scope}[shift={( 0, -7.5)}]
\foreach \y in {0, ..., 4}
{
    \draw (-1.5*1+\gw, \y) -- ( 1.5*1-\gw, \y);
}
\draw[dotted, fill=white] ( 1.5*0-\gw, 1.5-\gh) rectangle ( 1.5*0+\gw, 1.5+\gh);
\draw[fill=statecolor, rounded corners] (1.5*1-\gw, -0.5) rectangle ( 1.5*1+\gw, 4.5);
\node[rotate=90] at (1.5*1, 2) {$\ket{\psi^{(\ell)}_{\ell'-1}}$};
\draw ( 1.5*1+\gw, 2.3) -- ( 1.5*2-\gw, 2.3);
\draw ( 1.5*1+\gw, 2.1) -- ( 1.5*2-\gw, 2.1);
\draw ( 1.5*1+\gw, 1.9) -- ( 1.5*2-\gw, 1.9);
\draw ( 1.5*1+\gw, 1.7) -- ( 1.5*2-\gw, 1.7);
\draw[fill=statecolor, rounded corners] (-1.5*1-\gw, -0.5) rectangle (-1.5*1+\gw, 4.5);
\node[rotate=90] at (-1.5*1, 2) {$\ket{\phi_{n-\ell'}}$};
\node at ( 1.5*4.5, 2) {$= \partial^2\bra{\phi} C(G) \ket{\psi} / (\partial G_{\ell'} \partial G_{\ell})$};
\end{scope}
\end{tikzpicture}

%% file: gfx/gate_application_neighboring.tikz
\begin{tikzpicture}[scale=0.6, >=stealth]
\foreach \y in {0, ..., 2}
{
    \draw (-1.5*1+\gw, 1.5*\y+0.5) -- ( 1.5*1-\gw, 1.5*\y+0.5);
    \pgfmathsetmacro{\w}{int(2-\y)}
    \node[gray] at (-\gw-0.3, 1.5*\y+0.8) {\footnotesize $i_{\w}$};
}
\node at (-1.5*1.5, 2) {$\ket{\psi'} =$};
\node[gray] at ( \gw+0.3, 2+0.3) {\footnotesize $j$};
\draw[fill=gatecolor] (-\gw, 2-\gh) rectangle ( \gw, 2+\gh);
\draw[fill=statecolor, rounded corners] (1.5*1-\gw, -0.5) rectangle ( 1.5*1+\gw, 4.5);
\node at ( 1.5*1, 2) {$\ket{\psi}$};
\end{tikzpicture}

%% file: gfx/gate_application_general.tikz
\begin{tikzpicture}[scale=0.6, >=stealth]
\foreach \y in {0, ..., 4}
{
    \draw (-1.5*1+\gw, \y) -- ( 1.5*1-\gw, \y);
    \pgfmathsetmacro{\w}{int(4-\y)}
    \node[gray] at (-\gw-0.3, \y+0.3) {\footnotesize $i_{\w}$};
}
\node[gray] at ( \gw+0.3, 3+0.3) {\footnotesize $j_1$};
\node[gray] at ( \gw+0.3, 1+0.3) {\footnotesize $j_3$};
\node at (-1.5*1.5, 2) {$\ket{\psi'} =$};
\draw[fill=gatecolor] (-\gw, 2.5) -- (-\gw, 2.5+\gh) -- ( \gw, 2.5+\gh) -- ( \gw, 2.5);
\draw[fill=gatecolor] ( \gw, 1.5) -- ( \gw, 1.5-\gh) -- (-\gw, 1.5-\gh) -- (-\gw, 1.5);
\draw[fill=statecolor, rounded corners] (1.5*1-\gw, -0.5) rectangle ( 1.5*1+\gw, 4.5);
\node at ( 1.5*1, 2) {$\ket{\psi}$};
\end{tikzpicture}

%% file: gfx/hole_contraction.tikz
\begin{tikzpicture}[scale=0.6, >=stealth]
\foreach \y in {0, ..., 4}
{
    \draw (-1.5*1+\gw, \y) -- ( 1.5*1-\gw, \y);
    \pgfmathsetmacro{\w}{int(4-\y)}
    \node[gray] at (-\gw-0.3, \y+0.3) {\footnotesize $i_{\w}$};
}
\node[gray] at ( \gw+0.3, 3+0.3) {\footnotesize $j_1$};
\node[gray] at ( \gw+0.3, 1+0.3) {\footnotesize $j_3$};
\node at (-1.5*2, 2) {$\partial G =$};
\draw[dotted, fill=white] (-\gw, 2.5) -- (-\gw, 2.5+\gh) -- ( \gw, 2.5+\gh) -- ( \gw, 2.5);
\draw[dotted, fill=white] ( \gw, 1.5) -- ( \gw, 1.5-\gh) -- (-\gw, 1.5-\gh) -- (-\gw, 1.5);
\draw[fill=statecolor, rounded corners] ( 1.5*1-\gw, -0.5) rectangle ( 1.5*1+\gw, 4.5);
\node at ( 1.5*1, 2) {$\ket{\psi}$};
\draw[fill=statecolor, rounded corners] (-1.5*1-\gw, -0.5) rectangle (-1.5*1+\gw, 4.5);
\node at (-1.5*1, 2) {$\ket{\phi}$};
\end{tikzpicture}

%% file: gfx/brickwall_to_sequential.tikz
\begin{tikzpicture}[scale=0.6, >=stealth]
\begin{scope}
\foreach \y in {0, ..., 5}
{
    \draw (-1.5*0.5, \y) -- ( 1.5*2.5, \y);
}
\foreach \x in {0, 2}
{
    \foreach \y in {0, 2, 4}
    {
        \draw[fill=gatecolor] ( 1.5*\x-\gw, \y+0.5-\gh) rectangle ( 1.5*\x+\gw, \y+0.5+\gh);
        \pgfmathsetmacro{\z}{int(3-\x)};
        \node at ( 1.5*\x, \y+0.5) {$G_{\z}$};
    }
}
\foreach \x in {1}
{
    \foreach \y in {1, 3}
    {
        \draw[fill=gatecolor] ( 1.5*\x-\gw, \y+0.5-\gh) rectangle ( 1.5*\x+\gw, \y+0.5+\gh);
        \pgfmathsetmacro{\z}{int(\x+1)};
        \node at ( 1.5*\x, \y+0.5) {$G_{\z}$};
    }
    \draw[fill=gatecolor] ( 1.5*\x-\gw, 5.5) -- ( 1.5*\x-\gw, 5.5-\gh) -- ( 1.5*\x+\gw, 5.5-\gh) -- ( 1.5*\x+\gw, 5.5);
    \draw[fill=gatecolor] ( 1.5*\x-\gw,-0.5) -- ( 1.5*\x-\gw,-0.5+\gh) -- ( 1.5*\x+\gw,-0.5+\gh) -- ( 1.5*\x+\gw,-0.5);
    \draw[->] (1.5*\x, 5.75) -- (1.5*\x, 6) to[out=90, in=90] (1.5*\x+0.5, 6) -- (1.5*\x+0.5, 5.75);
    \draw[->] (1.5*\x+0.5,-0.75) -- (1.5*\x+0.5,-1) to[out=-90, in=-90] (1.5*\x,-1) -- (1.5*\x,-0.75);
}
\end{scope}
\draw[->, thick] ( 1.5, -1.75) -- ( 1.5, -2.5);
\begin{scope}[shift={(-1.5*3,-9)}]
\foreach \y in {0, ..., 5}
{
    \draw (-1.5*0.5, \y) -- ( 1.5*8.5, \y);
}
\foreach \x in {0, 2}
{
    \foreach \y in {0, 2, 4}
    {
        \pgfmathsetmacro{\s}{int(3*\x+0.5*\y)};
        \draw[fill=gatecolor] ( 1.5*\s-\gw, \y+0.5-\gh) rectangle ( 1.5*\s+\gw, \y+0.5+\gh);
        \pgfmathsetmacro{\z}{int(3-\x)};
        \pgfmathsetmacro{\w}{int(3-0.5*\y)};
        \node at ( 1.5*\s, \y+0.5) {$G_{\scalebox{.8}{$\scriptscriptstyle \z,\w$}}$};
    }
}
\foreach \x in {1}
{
    \foreach \y in {1, 3}
    {
        \pgfmathsetmacro{\s}{int(3*\x+0.5*(\y-1))};
        \draw[fill=gatecolor] ( 1.5*\s-\gw, \y+0.5-\gh) rectangle ( 1.5*\s+\gw, \y+0.5+\gh);
        \pgfmathsetmacro{\z}{int(\x+1)};
        \pgfmathsetmacro{\w}{int(3-0.5*(\y-1))};
        \node at ( 1.5*\s, \y+0.5) { $G_{\scalebox{.8}{$\scriptscriptstyle \z,\w$}}$};
    }
    \pgfmathsetmacro{\s}{int(3*\x+2)};
    \draw[fill=gatecolor] ( 1.5*\s-\gw, 5.5) -- ( 1.5*\s-\gw, 5.5-\gh) -- ( 1.5*\s+\gw, 5.5-\gh) -- ( 1.5*\s+\gw, 5.5);
    \draw[fill=gatecolor] ( 1.5*\s-\gw,-0.5) -- ( 1.5*\s-\gw,-0.5+\gh) -- ( 1.5*\s+\gw,-0.5+\gh) -- ( 1.5*\s+\gw,-0.5);
    \draw[->] (1.5*\s, 5.75) -- (1.5*\s, 6) to[out=90, in=90] (1.5*\s+0.5, 6) -- (1.5*\s+0.5, 5.75);
    \draw[->] (1.5*\s+0.5,-0.75) -- (1.5*\s+0.5,-1) to[out=-90, in=-90] (1.5*\s,-1) -- (1.5*\s,-0.75);
}
\end{scope}
\end{tikzpicture}

%% file: gfx/directed_gradient_graph.tikz
\begin{tikzpicture}[scale=0.6, >=stealth]
\begin{scope}[shift={(0,-6)}]
\draw[->] (-6.5, 3) to[out=0, in=90, looseness=0.8] (-5.75, 0.5);
\draw[->] (-3.5, 6) to[out=180, in=90, looseness=0.8] (-4.25, 3.5) -- (-4.25, 0.5);
\node at (-5, 0) {$\bra{\phi_0} Z_n \ket{\psi_{n-1}}$};
\node at (-2.5, 0) {\dots};
\draw[->] (-1.5, 3) to[out=0, in=90, looseness=0.8] (-0.75, 0.5);
\node at (0, 0) {$\bra{\phi_{n-\ell}} Z_{\ell} \ket{\psi_{\ell-1}}$};
\draw[->] (1.5, 6) to[out=180, in=90, looseness=0.8] (0.75, 3.5) -- (0.75, 0.5);
\node at (2.5, 0) {\dots};
\draw[->] (3.5, 3) to[out=0, in=90, looseness=0.8] (4.25, 0.5);
\node at (5, 0) {$\bra{\phi_{n-1}} Z_1 \ket{\psi_0}$};
\draw[->] (6.5, 6) to[out=180, in=90, looseness=0.8] (5.75, 3.5) -- (5.75, 0.5);
\end{scope}
\begin{scope}[shift={(0,0)}]
\draw[fill=gatecolor] (5-\gw,-\gh) rectangle (5+\gw, \gh);
\node at (5, 0) {$G_1$};
\draw[->] (6.5, 0) node[above] {$\ket{\psi_0}$} -- (5+\gw, 0);
\draw[->] (5-\gw, 0) -- node[above] {$\ket{\psi_1}$} (3.5, 0);
\node at (2.5, 0) {\dots};
\draw[fill=gatecolor] (-\gw,-\gh) rectangle ( \gw, \gh);
\node at (0, 0) {$G_{\ell}$};
\draw[->] (1.5, 0) node[above] {$\ket{\psi_{\ell-1}}$} -- (\gw, 0);
\node at (-2.5, 0) {\dots};
\draw[->] (-\gw, 0) -- node[above] {$\ket{\psi_{\ell}}$} (-1.5, 0);
\draw[->] (-3.5, 0) node[above] {$\ket{\psi_{n-1}}$} -- (-5+\gw, 0);
\end{scope}
\begin{scope}[shift={(0,-3)}]
\draw[fill=gatecolor] (-5-\gw,-\gh) rectangle (-5+\gw, \gh);
\node at (-5, 0) {$G_n$};
\draw[->] (-6.5, 0) node[above] {$\ket{\phi_0}$} -- (-5-\gw, 0);
\draw[white, ultra thick] (-4.5, 0) -- (-4, 0);
\draw[->] (-5+\gw, 0) -- (-3.5, 0) node[above] {$\ket{\phi_1}$};
\node at (-2.5, 0) {\dots};
\draw[fill=gatecolor] (-\gw,-\gh) rectangle ( \gw,\gh);
\node at (0, 0) {$G_{\ell}$};
\draw[->] (-1.5, 0) node[above] {$\ket{\phi_{n-\ell}}$} -- (-\gw, 0);
\draw[white, ultra thick] (0.5, 0) -- (1, 0);
\node[white, above] at (1.5, 0) {$\mathbf{\ket{\phi_{n-\ell+1}}}$};
\draw[->] (\gw, 0) -- (1.5, 0) node[above] {$\ket{\phi_{n-\ell+1}}$};
\node at (2.5, 0) {\dots};
\draw[->] (3.5, 0) -- node[above] {$\ket{\phi_{n-1}}$} (5-\gw, 0);
\end{scope}
\begin{scope}[shift={(0,-9)}]
\draw[->] (-5, 2.5) -- (-1.5, 0.65);
\draw[->] ( 0, 2.5) -- ( 0,   0.65);
\draw[->] ( 5, 2.5) -- ( 1.5, 0.65);
\node at (0, 0) {$\sum_{\ell=1}^n \Tr\big[Z_{\ell}^T \cdot \frac{\partial \bra{\phi_0} C(G) \ket{\psi_0}}{\partial G_{\ell}}\big]$};
\end{scope}
\end{tikzpicture}